\documentclass[a4paper]{article}

\usepackage{latexsym}
\usepackage[UKenglish]{babel}
\usepackage{graphicx,subfigure}
\usepackage{algorithm}
\usepackage{algpseudocode}
\usepackage{epstopdf}
	\graphicspath{{./}{figures/}}
\usepackage[colorlinks=true,linkcolor=blue,citecolor=blue]{hyperref}
\usepackage{xcolor}
\usepackage{amsfonts,amsmath,amsthm,mathtools,bbm,cool}

\newtheorem{theorem}{Theorem}[section]

\theoremstyle{remark}\newtheorem{remark}[theorem]{Remark}

\newcommand{\be}{\begin{equation}}
\newcommand{\ee}{\end{equation}}

\newcommand{\rev}{\textcolor{black}}

\allowdisplaybreaks

\begin{document}
\title{Kinetic models for epidemic dynamics in the presence of opinion polarization}

\author{Mattia Zanella \\
		{\small	Department of Mathematics ``F. Casorati''} \\
		{\small University of Pavia, Italy} \\
		{\small\tt mattia.zanella@unipv.it}} 
\date{}

\maketitle

\begin{abstract}
Understanding the impact of collective social phenomena in epidemic  dynamics is a crucial task to effectively contain the disease spread. In this work we build a mathematical description for assessing the interplay between opinion polarization and the evolution of a disease. The proposed kinetic approach describes the evolution of aggregate quantities  characterizing the agents belonging to epidemiologically relevant states, and will show that the spread of the disease is closely related to consensus dynamics distribution in which opinion polarization may emerge. In the present modelling framework, microscopic  consensus formation dynamics can be linked to macroscopic epidemic trends to trigger the collective adherence to protective measures. We conduct numerical investigations which confirm the ability of the model to describe different phenomena related to the  spread of an epidemic. 
\medskip


\end{abstract}

\tableofcontents

\section{Introduction}

During the outbreak of SARS-CoV-2 pandemic, we observed how, as cases escalated, collective compliance to the so-called non-pharmaceutical interventions (NPIs) was crucial to ensure public health in the absence of effective treatments, see e.g. \cite{APZ21,BC,BBDP,Gatto,Vig,Z}. Nevertheless, the effectiveness of lockdown measures heavily depended on the beliefs/opinions of individuals regarding protective behavior, which are thus linked to personal situational awareness \cite{DC,Tc}. Recent experimental results have shown that social norm changes are often triggered by opinion alignment  phenomena \cite{Tu}. In particular, the perceived adherence of individuals' social network has a strong impact on the effective support of the protective behaviour. The individual responses to threat is a core question to set-up effective measures prescribing norm changes in daily social contacts \cite{DFD} and cases escalation is a factor that may be perceived in different ways. For these reasons, it appears natural to couple classical epidemiological models with opinion dynamics in order to understand the mutual influence of these phenomena. 

In recent years the study of emerging properties of large systems of agents have obtained a growing interest in heterogeneous communities in social and life sciences, see e.g.  \cite{BCC,BDZ,CFRT,CFTV,CMPS,CPT,CPS,DM,FHT,HT,MT}. In particular, thanks to their cooperative nature, the dynamics leading to opinion formation phenomena have been often described through the methods of statistical mechanics \cite{BN,CFL,HK,SWS,W}. Amongst other approaches, kinetic theory provided a sound theoretical framework to investigate the emerging patterns of such systems \cite{DMPW,DW,T}. In this modelling setting, the microscopic, individual-based, opinion variations take place through binary interaction schemes involving the presence of social forces, whose effects are observable at the macroscopic scale \cite{PTTZ}. The equilibrium distribution describes the formation of a relative consensus about certain opinions \cite{PT13,T,TTZ}. In this direction, it is of paramount importance to obtain reduced complexity models whose equilibrium distribution is explicitly available under minimal assumptions \cite{FPTT,T}. The deviation from global consensus  appears in the form of opinion polarization, i.e. the divergence away from central positions towards extremes \cite{LRT}. This latter feature of the agents' opinion distribution is frequently observed in problems of choice formation \cite{ANT}. 

The derivation of classical compartmental epidemiological dynamics from particle systems have been recently explored as a follow-up question on the effectiveness of available modelling approaches. Indeed, epidemics, as well as many other collective phenomena, can be easily thought as a result of repeated interactions between a large number of individuals that eventually modify their epidemiological state. The transition rates between epidemiologically relevant states are furthermore influenced by several phenomena linked to the disease itself, and to the social behaviour of individuals.  Without attempting to revise the whole literature, we mention \cite{ABBDPTZ,BBDP,DMLT,DPaTz,DPTZ,DTZ,LT,LT} and the references therein for an introduction to the subject. Amongst them, contact dynamics are particularly relevant for contact-based disease transmissions. 

In this work we introduce a novel kinetic model that takes into account opinion formation dynamics of the individuals' protective behaviour coupled with epidemic spreading. These dynamics will result structurally linked due to the mutual influence of opinion formation processes and the transmission of the infection. \rev{The effects of behavioural dynamics on epidemic models has been investigated at the population level, see \cite{Poletti2009}. In particular, the formation of opinion clustering is connected to vaccination hesitancy,  see e.g. \cite{BDMOG} and the references therein. In this direction, we mention the recent results in \cite{DMLM,GFPS,KGFPS,ZZCS} where agent based dynamics are upscaled at the level of observable epidemiological quantities.
}

\rev{Kinetic equations are capable to provide efficient methods to bridge the microscopic, often unobservable, scale of individual agents, where elementary fundamental dynamics take place, and the macroscopic scale of observable manifestations. Indeed, in classical kinetic theory, the possibility to derive hydrodynamic descriptions of particles' systems is of paramount importance for provide real-time predictions. In the context of multiagent systems, the problem of deriving macroscopic equations is underexplored and has to face additional challenges in the definition of the social forces involved in the interactions. In order to get analytical insights on the macroscopic behaviour of the system the derivation of reduced complexity models is a key point. Hence, thanks to the derived surrogate models we can derive equilibrium profiles that are coherent with the ones defined at the kinetic level. In this work, we exploit the Fokker-Planck modelling approach that  has been introduced in \cite{T} for opinion formation processes. We remark that, at variance with \cite{DPTZ,DTZ}, the interactions between agents are structurally binary to mimic compromise behaviour. } The new derived \rev{macroscopic models encode} all the information of the opinion-based interactions, and describes coherent transition rates penalizing agents clustering on a weak protective behaviour. We will observe how opinion polarization can  trigger an increasing spread of  infection in society. 
 
In more details, the paper is organized as follows: in Section \ref{sect:kinetic} we introduce a kinetic epidemic model where agents are characterized by their epidemiological state and their opinion. Hence, a reduced complexity operator is derived to compute the large time opinion distribution of the system of agents and we discuss  minimal assumptions to observe opinion polarization. In Section \ref{sect:macro} we derive a macroscopic system of equations by considering an equilibrium closure method. The derived macroscopic model expresses the evolution at the epidemic scale  of the conserved quantities in the operator for opinion exchanges. Finally, in Section \ref{sect:num} we present several numerical tests showing the coherence of the presented closure strategy with the initial kinetic model in suitable scales. Furthermore, in the latter section we explore the possibility of considering more complex interaction functions in the opinion exchange process together with the influence of opinion polarization on the spreading of the disease. 

\section{A kinetic model approach for consensus formation and epidemic dynamics}\label{sect:kinetic}
In this section we introduce a kinetic compartmental model for the spreading of an infectious disease that is coupled with the evolution of the opinions' of individuals. We consider a system of agents that can be subdivided in the following epidemiologically relevant states: susceptible (S) agents are the ones that can contract the disease, infectious agents (I) are responsible for the spread of the disease, exposed (E) have been infected but are still not contagious and, finally, removed (R) agents cannot spread the disease. Each agent is endowed of a continuous opinion variable $w \in I$ which varies continuously in $I = [-1,1]$, where $-1$ and $1$ denote two opposite beliefs on the protective behaviour.  In particular, $w = -1$ means that the agents do not believe in the necessity of protections (like wearing masks or reducing daily contacts) whereas $w= 1$ is linked to maximal agreement on protective behaviour. We also assume that agents characterized by high protective behaviour are less likely to contract the infection. 

With the aim to incorporate the impact of opinion evolution in the dynamics of infection we denote by $f_J(w,t)$ the distribution of opinions at time $t \ge 0$ of agents in the compartment $J \in \mathcal C = \{S,E,I,R\}$. In particular, $f_J = f_J(w,t):[-1,1] \times \mathbb R_+ \rightarrow \mathbb R_+$ is such that $f_J(w,t)dw$ represents the fraction of agents with opinion in $[w,w+dw]$ at time $t\ge 0$ in the $J$th compartment. Furthermore, we impose
\[
\sum_{J \in \mathcal C} f_J(w,t) = f(w,t), \qquad  \int_{-1}^{1} f(w,t)dw = 1,
\]
while the mass fractions of the population in each compartment and their moment of order $r>0$ are given by 
\begin{equation}
\label{eq:rhom}
\rho_J(t) = \int_{-1}^1 f_J(w,t)dw, \qquad \rho_J(w,t) m_{r,J} =  \int_{-1}^1 \rev{w}^r f_J(w,t)dw. 
\end{equation}
In the following, to simplify notations, we will indicate with $m_J(t)$, $J \in \mathcal C$, the mean opinion in the compartment $J$ corresponding to $r = 1$. 

\begin{figure}
\centering
\includegraphics[scale = 0.3]{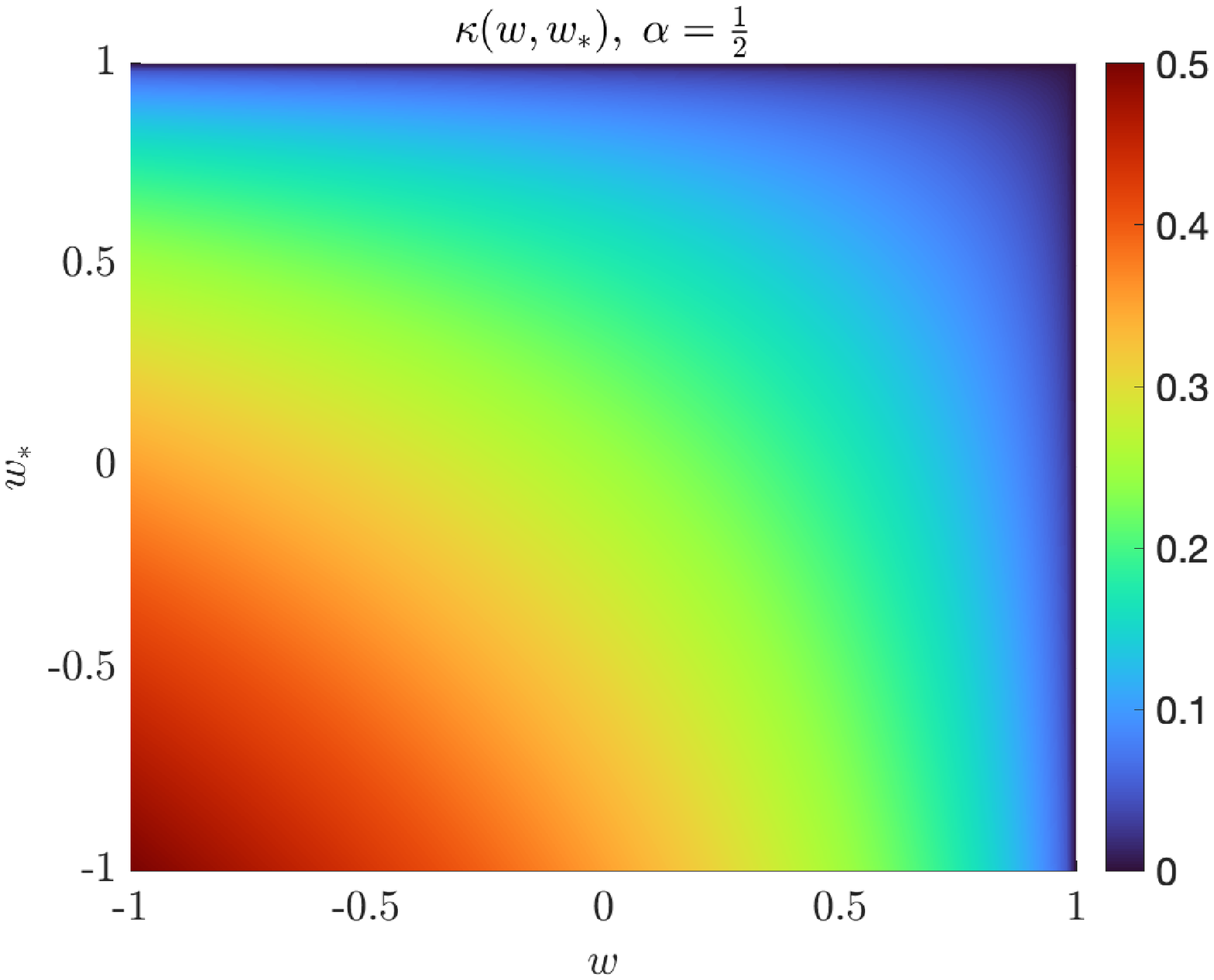}
\includegraphics[scale = 0.3]{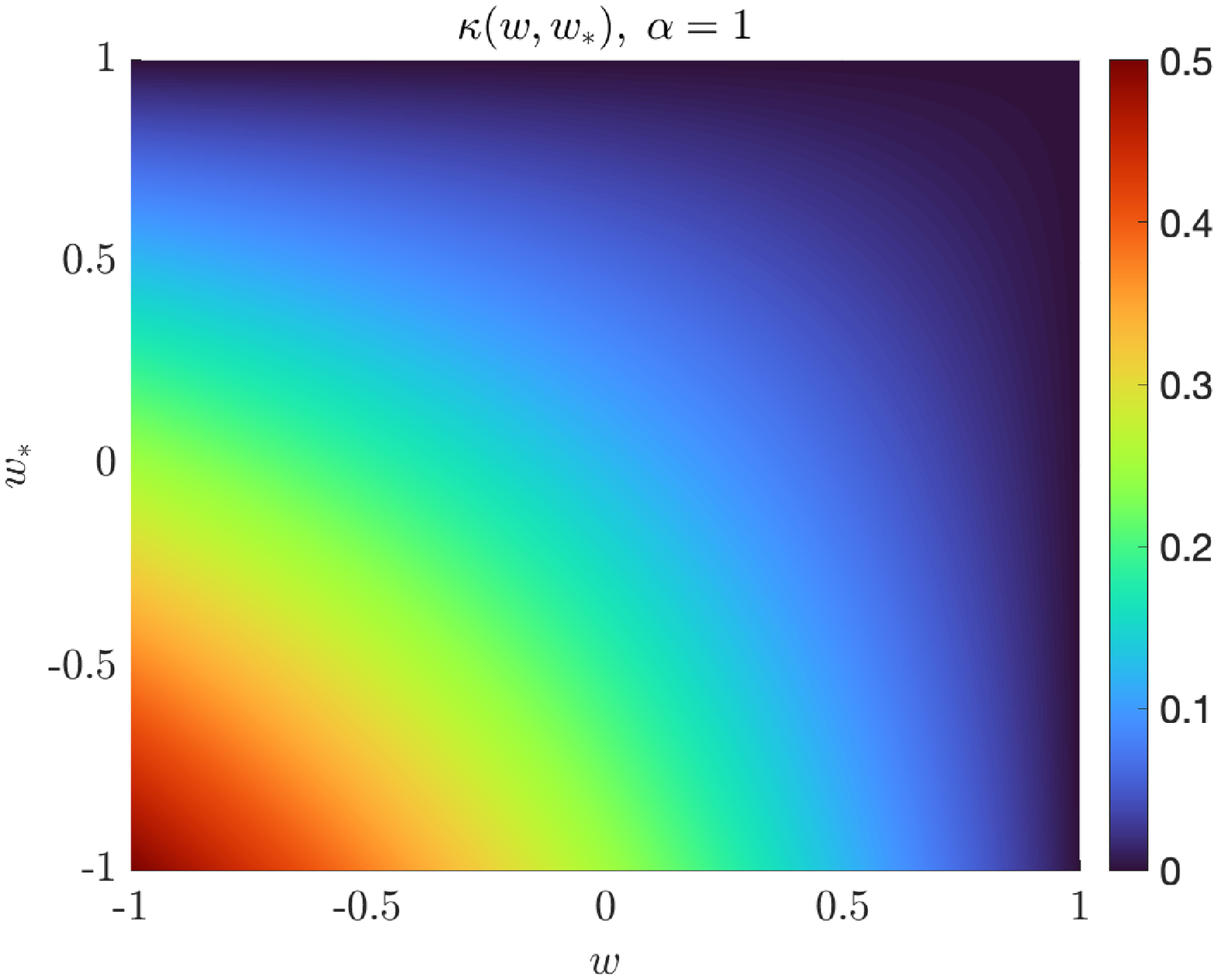}
\caption{We sketch the function $\kappa(w,w_*)$ in \eqref{eq:kappa_def} for $\alpha=\frac{1}{2}$ (left) and $\alpha = 1$ (right). In both cases, we fixed the coefficient $\beta = \frac{1}{2}$. }
\label{fig:kap}
\end{figure}

We assume that the introduced compartments of the model can have different impact in the opinion dynamics. The kinetic model for the coupled evolution of opinions and  infection is given by the following system of kinetic equations
\begin{equation}
\label{eq:kinetic_opinion}
\begin{split}
\partial_t f_S(w,t) &= -K(f_S,f_I)(w,t) + \dfrac{1}{\tau} Q_{S}(f_S,f_S)(w,t), \\
\partial_t f_E(w,t) &= K(f_S,f_I)(w,t) - \sigma_E f_E(w,t) + \dfrac{1}{\tau}Q_{E}(f_E,f_E)(w,t), \\
\partial_t f_I(w,t) &= \sigma_E f_E(w,t) - \gamma f_I(w,t) + \dfrac{1}{\tau} Q_{I}(f_I,f_I)(w,t), \\
\partial_t f_R(w,t) &= \gamma f_I(w,t) + \dfrac{1}{\tau}Q_{R}(f_R,f_R)(w,t),
\end{split}
\end{equation}
where $\tau>0$ and $Q_{J}(\cdot,\cdot)$ characterizes the evolution of opinions of agents that belong to the compartment $J \in \mathcal C$. In the next section we will specify the form of these operators describing binary opinion interactions among agents. The parameter $\sigma_E>0$ is such that $1/\sigma_E$ measures the mean latent period for the disease, whereas $\gamma>0$ is such that $1/\gamma>0$ is the mean infectious period \cite{DH}.  In \eqref{eq:kinetic_opinion} the transmission of the infection is governed by the local incidence rate 
\begin{equation}
\label{eq:K_def}
K(f_S,f_I)(w,t) =  f_S(w,t) \int_{-1}^1 \kappa(w,w_*)f_I(w_*,t)dw_*,
\end{equation}
where $\kappa(w,w_*)$ is a nonnegative decreasing function measuring the impact of the protective behaviour among different compartments. A leading example for the function $\kappa(w,w_*)$ can be obtained by assuming 
\begin{equation}
\label{eq:kappa_def}
\kappa(w,w_*) = \dfrac{\beta}{4^{\alpha}} (1-w)^{\alpha}(1-w_*)^{\alpha},
\end{equation}
where $\beta>0$ is the baseline transmission rate characterizing the epidemics and $\alpha>0$ is a coefficient linked to the efficacy of the protective measures. In Figure \ref{fig:kap} we represent the introduced function $\kappa(\cdot,\cdot)$ for several values of $\alpha>0$. \rev{We may observe how for $\alpha\equiv 0$ the influence of opinion dynamics on the epidemiological model disappears}. 
We highlight that in the simple case $\alpha = 1$ we get
\[
K(f_S,f_I)(w,t) = \dfrac{\beta}{4}(1-w)f_S(w,t)(1-m_I(t))I(t)\ge0, \qquad I(t)\ge 0
\]
with $K(f_S,f_I) \equiv 0$ in the case $m_I\equiv 1$ or in the case where all susceptible agents are concentrated in the maximal protective behaviour $w = 1$. 

\subsection{Kinetic models for opinion formation}
The dynamics of opinion formation have often been described by resorting to methods of statistical physics, see e.g. \cite{CFL,G}. In particular, kinetic theory provide a sound theoretical background to model fundamental interactions among agents and to provide a convenient dynamical structure for related follow-up questions on control problems and network formation \cite{APZ17,T}. In the aforementioned kinetic models, the opinion  variation of large systems of agents depends on binary interactions whose are driven by social forces determining the formation of consensus about certain opinions. The emerging distribution of opinions can be evaluated at the macroscopic level \cite{MT,PT13}. Recent advancements have been devoted to include external influences in opinion formation models to capture realistic complex phenomena. Without intending to review the very huge literature on the topic, we mention \cite{BN,CT,DMPW,DW} and the references therein. 

The elementary interactions between agents weight two opposite behaviour, the first is the compromise propensity, i.e. the tendency to reduce the opinion distance after interaction, and the second is the self-thinking, corresponding to unpredictable opinion deviations. In details, an interaction between two individuals in the compartments $J \in \mathcal C$ with opinion pair $(w,w_*)$ leads to an opinion pair $(w^{\prime}, w^{\prime}_*)$ defined by the relations
 \begin{equation}
 \label{eq:bin_int}
 \begin{split}
 w^\prime &= w + \lambda_J P(w,w_*) (w_*-w) + D(w)\eta_{J} \\
 w^\prime_* &= w_* + \lambda_J P(w_*,w)(w-w_*) + D(w_*)\tilde \eta_{J}, 
 \end{split}
 \end{equation}
 where $\lambda_J \in (0,1)$ and $P(w,w_*) \in [0,1]$ is an interaction function.  In \eqref{eq:bin_int} we further introduce the local diffusion function $D(w)$, and $\eta_{J},\tilde\eta_{J}$ are independent and identically distributed centered random variables with finite variance $ \left \langle \eta_{J} \right \rangle= \left \langle \eta_{J} \right \rangle= \sigma_J^2$, where we indicate with $\left\langle \cdot \right\rangle$  the expected value with respect to the distribution of the random variables. 

As observed in \cite{PTTZ} we have that the mean opinion is conserved for symmetric interaction functions, $P(w,w_*) = P(w_*,w)$ for all $w,w_* \in [-1,1]$. Indeed, from \eqref{eq:bin_int} we get 
\rev{
\begin{equation}
\label{eq:Pmean}
\left \langle w^\prime + w_*^\prime \right\rangle = w+w_* +\lambda_J( P (w,w_*) -  P(w_*,w) )(w_*-w),
\end{equation}
}
which reduces to $\left \langle w^\prime + w_*^\prime \right\rangle = w+w_*$ under the aforementioned assumptions. 
Furthermore, if we consider the mean energy we get
\[
\begin{split}
\left\langle (w^\prime)^2 +  (w^\prime_*)^2 \right\rangle =& w^2 + w_*^2 +\lambda_J^2\left[ P^2(w,w_*) +P^2(w_*,w)\right] (w_*-w)^2 \\
&+2\lambda_J[ P(w,w_*)w -  P(w_*,w)w_*](w_*-w)\\
&+ \sigma_J^2(D^2(w)+D^2(w_*)),
\end{split}
\]
meaning that the energy is not conserved on average in a single binary interaction. In the absence of the stochastic component, $\sigma_J^2\equiv 0$, we get that for symmetric interactions the mean energy is dissipated 
\[
\left\langle (w^\prime)^2 +  (w^\prime_{*})^2 \right\rangle = w^2 + w_*^2 - 2\lambda_J P(w,w_*)(w_*-w)^2 + o(\lambda_J)\le w^2 + w_*^2 + o(\lambda_J)
\]
The physical admissibility of interaction rules \eqref{eq:bin_int} is provided if $|w_{}^\prime|, |w_{*}^\prime| \le 1$ for $|w|,|w_*| \le 1$. We observe that 
\[
\begin{split}
|w_{}^\prime|& \le |(1-\lambda_JP(w,w_*))w + \lambda_J P(w,w_*)w_* + D(w)\eta_J| \\
&\le (1-\lambda_J P(w,w_*))|w|+\lambda_J P(w,w_*) + D(w)|\eta_J|,
\end{split}\]
since $|w_*| \le 1$, from which we get that the sufficient condition for $|w_{}^\prime| \le 1$ is provided by 
\[
D(w)|\eta_{J}| \le (1-\lambda_J P(w,w_*))(1-|w|),
\]
which is satisfied if a constant $c>0 $ exists and is such that 
\begin{equation}
\label{eq:cond_JH}
\begin{cases}
|\eta_{J}| \le c(1-\lambda_J P(w,w_*)) \\
c \cdot D(w) \le 1-|w|,
\end{cases}
\end{equation}
for all $w,w_* \in [-1,1]$. 
Since $0\le P(\cdot,\cdot)\le 1$ by assumption, the first condition in \eqref{eq:cond_JH} can be enforced by requiring that 
\[
|\eta_{J}| \le c(1-\lambda_J).
\]
Therefore it is sufficient to consider the support of the random variables determined by $|\eta_J| \le c(1-\lambda_J)$. The second condition in \eqref{eq:cond_JH}  forces $D(\pm 1)=0$. Other choices for the local diffusion function have been investigated in \cite{PTTZ,T}.\\

The collective trends of a system of agents undergoing  binary interactions \eqref{eq:bin_int} are determined by a Boltzmann-type model having the form 
\begin{equation}
\label{eq:kinetic_Qsum}
\partial_t f_J(w,t) = \dfrac{1}{\tau} Q_{J}(f_J,f_J),
\end{equation}
with $\tau>0$ and 
\[
Q_{J}(f_J,f_J)(w,t) = \left\langle\int_{-1}^1 \left(\dfrac{1}{{}^\prime \mathcal J} f_J({}^\prime w,t)f_J({}^\prime w_*,t)-f_J(w,t)f_J(w_*,t) \right)dw_* \right \rangle,
\]
where $({}^\prime w,{}^\prime w_*)$ are pre-interaction opinions generating the post-interaction opinions $(w,w_*)$ and ${}^\prime \mathcal J$ is the Jacobian of the transformation $({}^\prime w,{}^\prime w_*) \rightarrow (w,w_*)$. 

\subsection{Derivation of a Fokker-Planck model}\label{subsect:FP}
The  equilibrium distribuion of the kinetic model \eqref{eq:kinetic_Qsum} is very difficult to obtain analytically. For this reason, several reduced complexity models have been proposed. In this direction, a deeper insight on the equilibrium distribution of the kinetic model can be obtained by introducing a rescaling of both the interaction and diffusion parameters having roots in the so-called grazing collision limit of the classical Boltzmann equation \cite{Cerc,PT13}. The resulting model has the form of an aggregation-diffusion Fokker-Planck-type equation, encapsulating the information of microscopic dynamics. For the obtained surrogate model, the study of asymptotic properties is typically easier than the original kinetic model. 

We start by observing that we can conveniently express the operators $Q_{J}(\cdot,\cdot)$ in weak form. Let $\varphi(w)$ denote a test function, thus for $J \in \mathcal C$ we have
\[
\begin{split}
&\int_{-1}^1 \varphi(w)Q_{J}(f_J,f_J)(w,t)dw  \\
&\qquad =  \left\langle \int_{-1}^1 (\varphi(w^\prime) - \varphi(w)) f_J(w,t)f_J(w,t)dw_*\,dw \right\rangle, 
\end{split}\]
where $w^\prime$ is defined in \eqref{eq:bin_int}. The prototype of a symmetric interaction function $P$ is given by the constant function $P\equiv 1$. In this case, we may obtain analytic insight on the large time distribution of the system by resorting to a reduced complexity Fokker-Planck-type model \cite{T}. We introduce the so-called quasi-invariant regime 
\begin{equation}
\label{eq:quasi}
 \lambda_J \rightarrow \epsilon \lambda_J,\qquad \sigma_J^2 \rightarrow \epsilon \sigma_J^2,
\end{equation}
where $\epsilon >0$ is a scaling coefficient. We have
\[
\begin{split}
&\varphi(w^\prime) - \varphi(w)  \\
&\quad= \varphi^\prime(w)\left\langle w^\prime-w \right\rangle + \dfrac{1}{2}\varphi^{\prime\prime}(w) \left\langle (w^\prime_{}-w)^2 \right\rangle  + \dfrac{1}{6}\varphi^{\prime\prime\prime}(\bar w_{})\left\langle (w^\prime_{}-w)^3 \right\rangle, 
\end{split}
\]
where $\min\{w,w_{}^\prime\}< \bar w_{}< \max\{w,w_{}^\prime\}$. 
Plugging the above expansions in the Boltzmann-type model we have
\begin{equation}
\label{eq:pass_bol}
\begin{split}
&\dfrac{d}{dt} \int_{-1}^1 \varphi(w)f_J(w,t)dw = \\
&\qquad \epsilon\lambda_J \rho_J \int_{-1}^1 \int_{-1}^1\varphi^\prime(w)(m_J-w)f_J(w,t)dw \\
&\qquad+ \dfrac{\epsilon\sigma^2}{2}\int_{-1}^1 \varphi^{\prime\prime}(w)D^2(w)f_J(w,t)dw + R(f_J,f_J),
\end{split}
\end{equation}
where $R(f_J,f_J)$ is a reminder term
\[
\begin{split}
&R(f_J,f_J)(w,t) = \dfrac{1}{2} \int_{-1}^{1}\varphi^{\prime\prime}(x)\epsilon^2\lambda^2_J(w_*-w)^2 f_J(w,t)dw  \\
&\quad+ \dfrac{1}{6}\left\langle \int_{-1}^1\int_{-1}^1 \varphi^{\prime\prime\prime}(w)(\epsilon\lambda_J(w_*-w) + D(w)\eta_{J})^3 f_J(w,t)f_J(w_*,t)dw\,dw_*\right\rangle
\end{split}
\]
Hence, in the time scale $\xi = \epsilon t$, introducing the distribution $g_J(w,\rev{\xi}) = f_J(w,\xi/\epsilon)$, we have that $\partial_\xi g_J(w,\xi) = \frac{1}{\epsilon} \partial_t f_J$  and \eqref{eq:pass_bol} becomes
\[
\begin{split}
&\dfrac{d}{d\xi}\int_{-1}^1 \varphi(w)g_J(w,\xi)dw =  \lambda_J  \int_{-1}^1 \int_{-1}^1 \varphi^\prime(w)(m_J-w)g_J(w,\xi)dw \\
&\qquad+ \dfrac{\sigma_J^2}{2}\int_{-1}^1\varphi^{\prime\prime}(w)D^2(w)g_J(w,\xi)dw + \dfrac{1}{\epsilon}R(g_J,g_J)(w,\xi),
\end{split}
\]
where now $\frac{1}{\epsilon }R(g_J,g_J) \rightarrow 0$ under the additional hypothesis $\left\langle |\eta_{J}|^3 \right\rangle<+\infty$, see \cite{CPT,T}. Consequently, for $\epsilon\rightarrow 0^+$, from the above equation we have 
\[
\begin{split}
\dfrac{d}{d\xi}\int_{-1}^1 \varphi(w)g_J(w,\xi)dw =&  \lambda_J \int_{-1}^1 \int_{-1}^1 \varphi^\prime(w)(m_J-w)g_J(w,\xi)dw \\
&+ \dfrac{\sigma_J^2}{2}\int_{-1}^1\varphi^{\prime\prime}D^2(w)g_J(w,\xi)dw. 
\end{split}\] 
Now, with a slight abuse of notation, we restore $t\ge 0$ as time variable \rev{and $f_J$ as distribution}. In view of the smoothness of $\varphi$, integrating back by parts the terms on the right hand side, we finally get the Fokker-Planck-type model 
\begin{equation}
\label{eq:FP}
\begin{split}
\partial_t f_J(w,t) &= \bar Q_J(f_J,\rev{f_J})(w,t) \\
&=  \partial_w \left[ \lambda_J(w- m_J)f_J(w,t) + \dfrac{\sigma_J^2}{2}\partial_w (D^2(w)f_J(w,t)) \right] 
\end{split}
\end{equation}
complemented by the following no-flux boundary conditions
\[
\begin{split}
\lambda_J(w-m_J)f_J(w,t) + \dfrac{\sigma_J^2}{2}\partial_w (D^2(w)f_J(w,t)) \Big|_{w = \pm 1} = 0 \\
D^2(w)f_J(w,t) \Big|_{w = \pm 1 } = 0.
\end{split}\]

We can observe that the steady state of the Fokker-Planck-type model \eqref{eq:FP} is analytically computable under suitable hypotheses on the local diffusion function. If $D(w) = \sqrt{1-w^2}$, then the large time behavior of the model is given by  a Beta distribution having the form
\begin{equation}
\label{eq:beta}
f_J^\infty(w) = \dfrac{(1+w)^{\frac{1+m_J}{\nu_J}-1} (1-w)^{\frac{1- m_J}{\nu_J}-1}}{2^{\frac{2}{\nu_J}-1}B\left(\frac{1+ m_J}{\nu_J},\frac{1- m_J}{\nu_J} \right)}, \quad \nu_J =\dfrac{ \sigma_J^2}{\lambda_J}, 
\end{equation}
where $B(\cdot,\cdot)$ indicates the Beta function. It is worth to highlight that the first two moments of the obtained Beta distribution are defined as follows
\begin{equation}
\label{eq:mom}
\int_{-1}^1 w f_J^\infty(w)dw = m_J;\qquad \int_{-1}^1 w^2 f^\infty_J(w)dw = \dfrac{\nu_J}{2+\nu_J} + \dfrac{2}{2+\nu_J} m_J^2.
\end{equation}

\begin{figure}
\centering
\includegraphics[scale = 0.3]{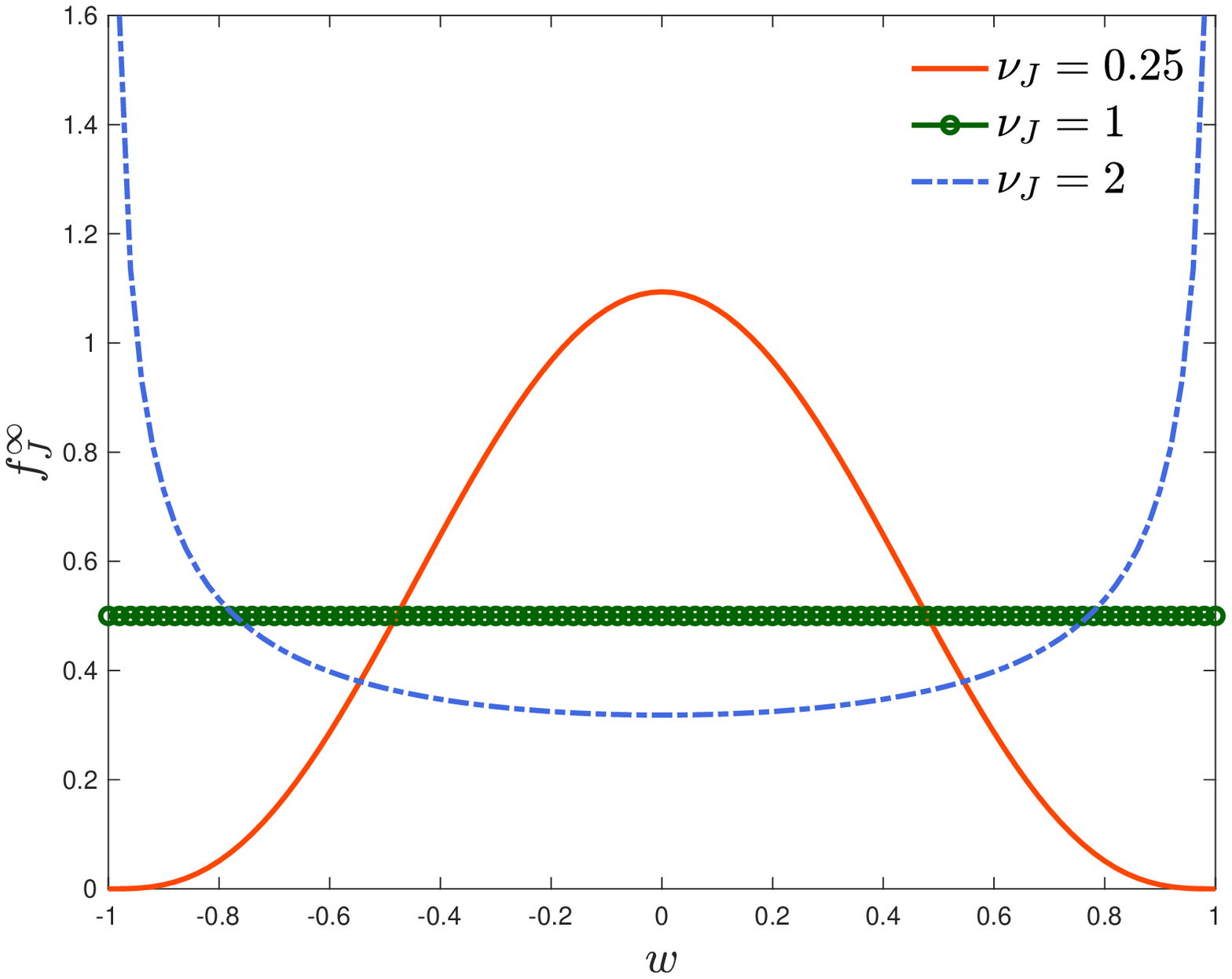}
\includegraphics[scale = 0.3]{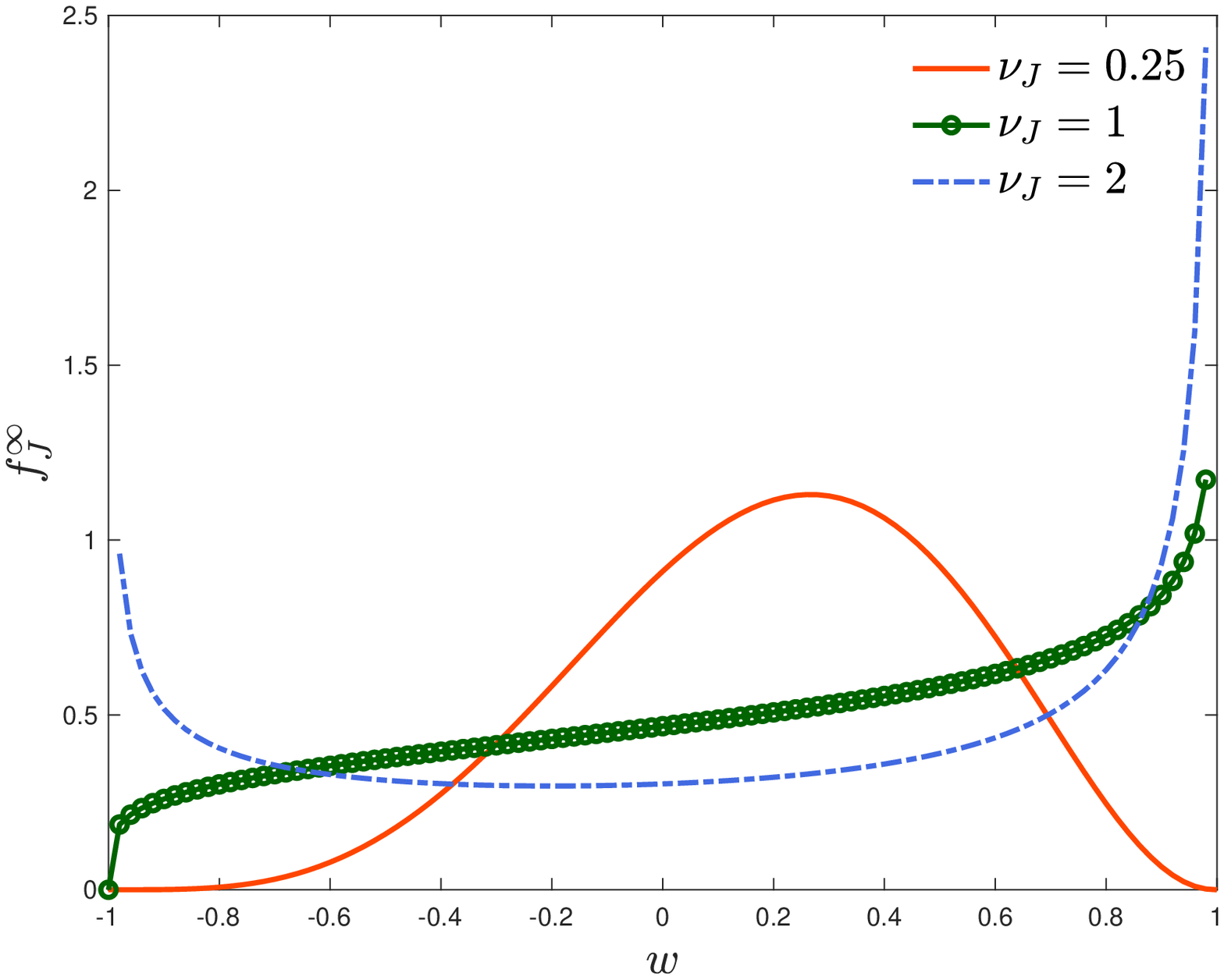}
\caption{We depict the equilibrium distribution \eqref{eq:beta} for several choices of the parameter $\nu_J>0$ and for $ m_J = 0$ (left) or $ m_J = 0.2$ (right). Opinion polarization is observed for $\nu_J>1$ whereas consensus formation corresponds to $\nu_J\ll 1$. }
\label{fig:pola}
\end{figure}

We can observe that the obtained model is suitable to describe classical consensus-type dynamics. This behaviour is observed if the compromise force is stronger than the one characterizing self-thinking, i.e. $\sigma_J^2<\lambda_J$. On the other hand, if self-thinking is stronger than the compromise propensity, i.e. $\sigma_J^2>\lambda_J$, we observe opinion polarization of the society. In Figure \ref{fig:pola} we depict the equilibrium distribution \eqref{eq:beta} for several choices of the parameter $\nu_J>0$. In the right figure we assume that $ m_J = 0$ whereas, in the left figure, we consider the asymmetric case with $ m_J = 0.2$. We may observe that opinion polarization is obtained in the case $\nu_J>1$ as discussed. 

\begin{remark}\label{rem:FP}
In the more general case where interactions between agents is weighted by a nonconstant function $P(w,w_*) \in [0,1]$, we may obtain the nonlocal Fokker-Planck-type model 
\[
\partial_t f_J(w,t) = \partial_w \left[ \mathcal B[f_J](w,t) f_J(w,t) + \dfrac{\sigma^2}{2}\partial_w f_J(w,t)\right]
\]
where
\[
\mathcal B[f_J](w,t)= \int_{-1}^1 P(w,w_*)(w-w_*)f_J(w_*,t)dw_*.
\]
In this case, it is difficult to get an analytical formulation of the steady state distribution. 
\end{remark}
\section{Macroscopic opinion-based SEIR dynamics}\label{sect:macro}
Once the equilibrium distribution of the operators \rev{$\bar Q_J(f_J,f_J)(w,t)$} is characterised, we can study the behaviour of the original system \eqref{eq:kinetic_opinion}. In this section we compute the evolution of observable macroscopic equations of the introduced kinetic model for epidemic dynamics with opinion-based incidence rate. 

\subsection{Derivation of moment based systems}
Let us rewrite the original model \eqref{eq:kinetic_opinion} with the reduced complexity Fokker-Planck-type operators defined in Section \ref{subsect:FP}. We obtain the following model
\begin{equation}
\label{eq:kin_2}
\begin{split}
\partial_t f_S(w,t) &= -K(f_S,f_I) + \dfrac{1}{\tau}\bar Q_{S}(f_S,\rev{f_S})(w,t),  \\
\partial_t f_E(w,t) &= K(f_S,f_I) -\sigma_E f_E(w,t) + \dfrac{1}{\tau}\bar Q_E(f_E,\rev{f_E})(w,t), \\
\partial_t f_I(w,t) &= \sigma_E f_E(w,t)-\gamma f_I(w,t) + \dfrac{1}{\tau}\bar Q_{I}(f_I,\rev{f_I})(w,t),  \\
\partial_t f_R(w,t ) &= \gamma f_I(w,t) + \dfrac{1}{\tau}\bar Q_{R}(f_R,\rev{f_R})(w,t)
\end{split}
\end{equation}
where $K(\cdot,\cdot)$ has been defined in \eqref{eq:K_def} \rev{and the collision-like operators $\bar Q_J(\cdot,\cdot)$, $J\in \mathcal C$, have been derived in Section \ref{subsect:FP}}. The system of kinetic equations \eqref{eq:kin_2} is further complemented by no-flux boundary conditions at $w = \pm 1$ and contains the information on the spreading of the epidemic in terms of the distribution of opinions of a population of agents. 

Integrating the model \eqref{eq:kinetic_opinion} with respect to the $w$ variable and recalling that, \rev{if the interaction function is symmetric},   the Fokker-Planck operators are mass and momentum preserving in the presence of no-flux boundary conditions \rev{coherently with what we observed for the microscopic binary scheme \eqref{eq:Pmean}}. \rev{Hence,} we obtain the evolution of mass fractions $\rho_J$, $J \in \mathcal C$, 
\begin{equation}
\label{eq:mass}
\begin{split}
\dfrac{d}{dt} \rho_S(t) &= - \dfrac{\beta}{4} \left(1-m_I - m_S + m_Sm_I\right)\rho_S\rho_I, \\
\dfrac{d}{dt} \rho_E(t) &=  \dfrac{\beta}{4} \left(1-m_I - m_S + m_Sm_I\right)\rho_S\rho_I - \sigma_E \rho_E, \\
\dfrac{d}{dt} \rho_I (t)&= \sigma_E \rho_E - \gamma \rho_I, \\
\dfrac{d}{dt} \rho_R(t) &= \gamma \rho_I,
\end{split}
\end{equation}
\rev{where we} observe that $(1-m_I-m_S + m_Sm_I )\rho_S\rho_I= (1-m_I)(1-m_S)\rho_S\rho_I\ge 0$ since $\rho_Im_I,\rho_Sm_S \in [-1,1]$. Unlike the classical SEIR model, the system for the evolution of mass fractions in \eqref{eq:mass} is not closed since  the evolution of $\rho_J$, $\rho_J \in \mathcal C$ depends on the evolution of the local mean opinions $m_J$, $J \in \mathcal C$. The closure of system \eqref{eq:mass} may be formally obtained by resorting to a limit procedure. The main idea is to observe that   the typical time scale of the opinion dynamics is faster than the one of the epidemic, and therefore $\tau\ll 1$. Consequently, for small values of $\tau$ the opinion distribution of the $J$th compartment reaches its local Beta-type equilibrium with a mass fraction $\rho_J$ and local mean opinion $m_J$ as verified in Section \ref{subsect:FP}. 
In particular, we observe exponential convergence of the derived Fokker-Planck equation \eqref{eq:FP} towards the local Maxwellian  parametrised by the conserved quantities, i.e. $\rho_J$ and $m_J$, see \cite{FPTT}. \rev{We highlight that this assumption is coherent with what stated in the work \cite{Poletti2009} since epidemic transmission is generally slower than the propagation of  information.  }

Hence, to get the evolution of mean values we can multiply by $w$ and integrate \eqref{eq:kin_2} to get system
\[
\begin{split}
\dfrac{d}{dt}(\rho_S(t)m_S(t)) &= - \dfrac{\beta}{4}\rho_I(1-m_I)\int_{-1}^1 w(1-w)f_S(w,t)dw, \\
\dfrac{d}{dt}(\rho_E(t)m_E(t)) &= \dfrac{\beta}{4}\rho_I(1-m_I)\int_{-1}^1 w(1-w)f_S(w,t)dw - \sigma_E m_E\rho_E, \\
\dfrac{d}{dt}(\rho_I(t)m_I(t)) &= \sigma_E m_E\rho_E - \gamma m_I\rho_I, \\
\dfrac{d}{dt}(\rho_R(t)m_R(t)) &= \gamma m_I\rho_I,
\end{split}
\]
which now depends on the second order moment, making this system not closed. It is now possible to close this expression by using the energy of the Beta-type local equilibrium distribution as in \eqref{eq:mom}. We have
\begin{equation}
\label{eq:m2}
m_{2,J} =\rho_J \dfrac{\nu_J + 2m_J^2}{2+\nu_J } ,
\end{equation}
%
where $\nu_S = \sigma^2/\lambda_S$ and $m_J$ is the local mean opinion in the $J$th compartment \eqref{eq:rhom} 

Hence, we have
\[
\dfrac{d}{dt}(\rho_S(t)m_S(t)) = -\dfrac{\beta}{4}(1-m_I)\rho_I\rho_S\left( m_S - \dfrac{\nu_S + 2m_S^2}{2+\nu_S} \right)
\]
which gives 
\[
\rho_S(t)\dfrac{d}{dt}m_S(t) = -\dfrac{\beta}{4}(1-m_I)\rho_I\rho_S\left( m_S - \dfrac{\nu_S + 2m_S^2}{2+\nu_S} \right) - m_S\dfrac{d}{dt}\rho_S
\]
where the time evolution of the fraction $\rho_S$ has been derived in the first equation of \eqref{eq:mass}. The evolution of the local mean $m_S$ is therefore given by 
\[
\dfrac{d}{dt}m_S(t) = \dfrac{\beta}{4}(1-m_I)\rho_I \left[ \dfrac{\nu_S+2m_S^2}{2+\nu_S} - m_S^2 \right]. 
\]
We may apply an analogous procedure for the remaining local mean values in the compartments of exposed, infected and recovered. to obtain 
\begin{equation}
\label{eq:mean}
\begin{split}
\dfrac{d}{dt} m_S(t) &= \dfrac{\beta}{4}\dfrac{\nu_S}{2+\nu_S}(1-m_I)\rho_I \left[ 1 - m_S^2 \right]. \\
\dfrac{d}{dt}m_E(t) &= \dfrac{\beta}{4}\dfrac{\rho_S\rho_I}{\rho_E}(1-m_I)\left[ m_S -  \left(\dfrac{\nu_S+2m_S^2}{2+\nu_S}  \right) - m_E(1-m_S) \right]  \\
\dfrac{d}{dt} m_I(t) &= \sigma_E \dfrac{\rho_E}{\rho_I} \left( m_E-m_I\right) \\
\dfrac{d}{dt}m_R(t) &= \gamma \dfrac{\rho_I}{\rho_R}\left(m_I - m_R\right).
\end{split}
\end{equation}

\begin{remark}
In the case of consensus of the susceptible agents, i.e. for $\nu_S \rightarrow 0^+$, we can observe that  $\frac{d}{dt}m_S(t) = 0$ which leads $m_S(t) = m_S(0)$ for all $t \ge 0$. The spread of the infection therefore depends only on the protective behavior of the agents on the compartment $I \in \mathcal C$. Furthermore, the trajectory of the second equation is decreasing in time since 
\[
\dfrac{d}{dt} m_E(t) = -\dfrac{\beta}{4}(1-m_I)\rho_I (1-m_S)\rho_S \dfrac{m_E}{\rho_E},
\]
and $\dfrac{\beta}{4}(1-m_I)\rho_I (1-m_S)\rho_S/\rho_E\ge 0$. 

\end{remark}

\begin{remark}
If the local incidence rate $K(f_S,f_I)$ in \eqref{eq:K_def}is such that $\kappa(w,w_*)\equiv \beta>0$ than  we easily observe that the evolution of mass fractions are decoupled with the local mean opinions since in this case integrating \eqref{eq:kinetic_opinion} we get 
\[
\begin{split}
\dfrac{d}{dt} \int_{-1}^1 f_S(w,t)dw &= -\beta \int_{-1}^1 f_S(w,t)dw \int_{-1}^1 f_I(w,t)dw, \\
\dfrac{d}{dt} \int_{-1}^1 f_E(w,t)dw &= \beta \int_{-1}^1 f_S(w,t)dw \int_{-1}^1 f_I(w,t)dw - \sigma_E \int_{-1}^1 f_E(w,t)dw,\\
\dfrac{d}{dt} \int_{-1}^1 f_I(w,t)dw &= \sigma_E \int_{-1}^1 f_E(w,t)dw - \gamma \int_{-1}^1 f_I(w,t)dw, \\
\dfrac{d}{dt} \int_{-1}^1 f_R(w,t)dw &= \gamma \int_{-1}^1 f_I(w,t)dw. 
\end{split}
\]
Therefore, the model \eqref{eq:kinetic_opinion} for constant $\kappa(w,w_*)\equiv \beta$ reduces to the  classical SEIR compartmental model.
\end{remark}

\rev{
\begin{remark}
In the case of non-symmetric interaction function $P(w,w_*)$ the system of macroscopic equations looses the information on the evolution of the mean values. A possible prototype of non-symmetric $P$ proposed in \cite{PTTZ} is the linear perturbation of a constant, i.e. $P(w,w_*) = P(w_*) = pw_*+q$, $q \in [0,1]$ and $|p| \le \min\{q,1-q\}$. In this case, in \cite{PTTZ} it is shown that the mean opinion is not conserved and that the asymptotic distribution functions are given by a Dirac delta distribution $\delta(w-1)$ if $p>0$ or by a Dirac delta $\delta(w+1)$ if $p<0$.  
\end{remark}
}

\subsection{The macroscopic model with saturated incidence rate}

It is not restrictive to suppose that infected agents possess enforced situational awareness. For this reasons, we may consider the case in which $m_I(t)=\bar{m}_I\in(0,1)$. From the first equation of \eqref{eq:mean} we get 
\[
\dfrac{d}{dt}m_S(t) = \dfrac{\beta}{4}\rho_I(t)(1-\bar m_I)\dfrac{\nu_S }{2+\nu_S}   \left[ 1- m_S^2(t)\right]
\]
with initial condition $m_S(0) = m_S^0 \in [-1,1]$. In particular, if $m_S^0 = \pm 1$ then $m_S(t) = m_S^0$ for all $t\ge 0$, otherwise if $-1<m_S^0<1$ we get 
\begin{equation}
\label{eq:mS_sat}
m_S(t) = \dfrac{\exp\{2 \int_0^t J(\rho_I(s))ds\} - \exp\{C_0\}}{\exp\{C_0\}  + \exp\{2 \int_0^t J(\rho_I(s))ds\}},
\end{equation}
with $C_0 = \log \frac{1-m_S^0}{1+m_S^0}$ and $J(\rho_I(s)) = \frac{\beta}{4}  \frac{\nu_S}{2+\nu_S}(1-\bar m_I)\rho_I(s)\ge 0$. We may easily observe that from \eqref{eq:mS_sat} we have $m_S(t) \in (-1,1)$ for all $t\ge 0$. 

Hence, plugging \eqref{eq:mS_sat} into the system for the mass fractions \eqref{eq:mass} we get
\begin{equation}
\label{eq:saturated}
\begin{split}
\dfrac{d}{dt}\rho_S(t) &= -\bar\beta H(t,\rho_I)\rho_S(t)\rho_I(t),  \\
\dfrac{d}{dt}\rho_E(t) &= \bar\beta H(t,\rho_I)\rho_S(t)\rho_I(t) - \sigma_E\rho_E,\\
\dfrac{d}{dt}\rho_I(t) &= \sigma_E\rho_E - \gamma\rho_I,\\
\dfrac{d}{dt}\rho_R(t) &= \gamma\rho_I
\end{split}
\end{equation}
where 
\[
\bar \beta H(t,\rho_I) = \bar \beta\left(1-\dfrac{e^{2\int_0^tJ(\rho_I(s))ds}-e^{C_0}}{e^{2\int_0^tJ(\rho_I(s))ds}+e^{C_0}}\right) \in (0,1), 
\]
and $\bar\beta = \frac{\beta}{4} (1-\bar m_I)$. In this case, model \eqref{eq:saturated} is a generalization of classical models with saturated incidence rate, see \cite{CS,KM}. In this setting, we derive the basic reproduction number by defining 
\[
D(\rho_S,\rho_I) = \bar \beta H(t,\rho_I)\rho_S\rho_I, 
\]
and the function $D(\rho_S,\rho_I)$ is such that 
\[
\dfrac{\partial D(\rho_S,\rho_I)}{\partial \rho_S} >0, \qquad \dfrac{\partial D(\rho_S,\rho_I)}{\partial \rho_I} >0
\]
and $D(\rho_S,\rho_I)$ is concave since $\frac{\partial^2}{\partial \rho_I^2}D(\rho_S,\rho_I)\le 0$ for all $\rho_S,\rho_I>0$. Hence, the basic reproduction number $R_0$ of the model is given by 
\[
R_0 = \dfrac{1}{\gamma}\lim_{\rho_I\to 0, \rho_S\to 1} \dfrac{\partial D(\rho_S,\rho_I)}{\partial \rho_I} = \dfrac{ \beta (1-\bar m_I)}{4\gamma}. 
\]
\rev{For the computation of the basic reproduction number $R_0$ using the method of next generation matrix we refer to \cite{BC}. The method goes back to \cite{Die} and we also refer to \cite{Die2} for an application to the SEIR model.}

\section{Numerical examples}\label{sect:num}

In this section we present several numerical examples to show the consistency of the proposed approach. \rev{Furthermore, we will show the impact of opinion consensus dynamics on observable epidemic quantities based on Beta-type equilibrium and on the macroscopic models generated by bounded-confidence-type opinion distributions. The consensus of the population on the adoption of protective measures is capable to reduce the epidemic peak together with the total number of infected agents. Finally, we will investigate numerically the impact of opinion polarization on the defined dynamics. }

\rev{From the methodological point of view, we will consider} classical direct simulation Monte Carlo (DSMC) methods \rev{to} show how, in the quasi-invariant limit defined in \eqref{eq:quasi}, the large time distribution of the Boltzmann-type model \eqref{eq:kinetic_Qsum} is consistent with the one obtained from the  reduced complexity Fokker-Planck model \eqref{eq:FP}. In the following, we will first concentrate on the case of interactions leading to a Beta distribution of the form \eqref{eq:beta}. As a follow-up question we will explore the observable effects of nonlinear interaction functions.  

Hence, in order to approximate the dynamics of the kinetic SEIR model  \eqref{eq:kinetic_opinion} \rev{for small values of $\tau>0$}, we resort to classical strong stability preserving schemes combined to recently developed semi-implicit structure preserving schemes for nonlinear Fokker-Planck equations \cite{PZ}, see also \cite{LZ} for further applications. These methods are capable to reproduce large time statistical properties of the exact steady state with arbitrary accuracy together with the preservation of the main physical properties of the solution, like positivity and entropy dissipation. \rev{Indeed, we highlight how in the present setting the development of DSMC methods would encounter severe time step restrictions depending on the values of $\tau>0$. We point the interested reader to \cite{PR} for a more detailed discussion on the topic. }

\subsection{Test 1: large time behaviour of kinetic opinion formation models}
In this section we test the consistency of the quasi-invariant limit to obtain a reduced complexity Fokker-Planck model. In particular, we concentrate on a kinetic model for opinion formation where the binary scheme is given by \eqref{eq:bin_int} in the simplified case $P\equiv 1$ and for $D(w) = \sqrt{1-w^2}$. As discussed in Section \ref{subsect:FP}, for quasi-invariant interactions as in \eqref{eq:quasi} and in the limit $\epsilon \rightarrow 0^+$, the emerging distribution can be computed through the Fokker-Planck model \eqref{eq:FP} and is given be the Beta distribution \eqref{eq:beta}. 

We rewrite the Boltzmann-type model \eqref{eq:kinetic_Qsum} as follows
\[
\partial_t f_J(w,t) = \dfrac{1}{\tau}\left(Q^+(\rev{f_J},\rev{f_J})(w,t) - \rev{f_J}(w,t)\right),
\]
where $\tau>0$ is a positive constant and
\[
Q^+(\rev{f_J},\rev{f_J})(w,t) = \left\langle \int_{-1}^1 \dfrac{1}{{}^\prime \mathcal J} \rev{f_J}({}^\prime w,t)\rev{f_J}({}^\prime w_*,t) dw_* \right\rangle, 
\]
where $({}^\prime w, {}^\prime w_*)$ are the pre-interaction opinions generating the post-interaction opinions $(w,w_*)$ according to the binary interaction rule \eqref{eq:bin_int} and ${}^\prime \mathcal J$ is the Jacobian of the transformation $({}^\prime w, {}^\prime w_*)\to (w,w_*)$. To compute the large time numerical solution of the introduced Boltzmann- type model we consider $N = 10^6$ particles and  we assume that $\tau = 1$. The quasi-invariant regime of parameters in \eqref{eq:quasi} is considered for $\epsilon = 10^{-1},10^{-3}$.  

\begin{figure}
\includegraphics[scale = 0.3]{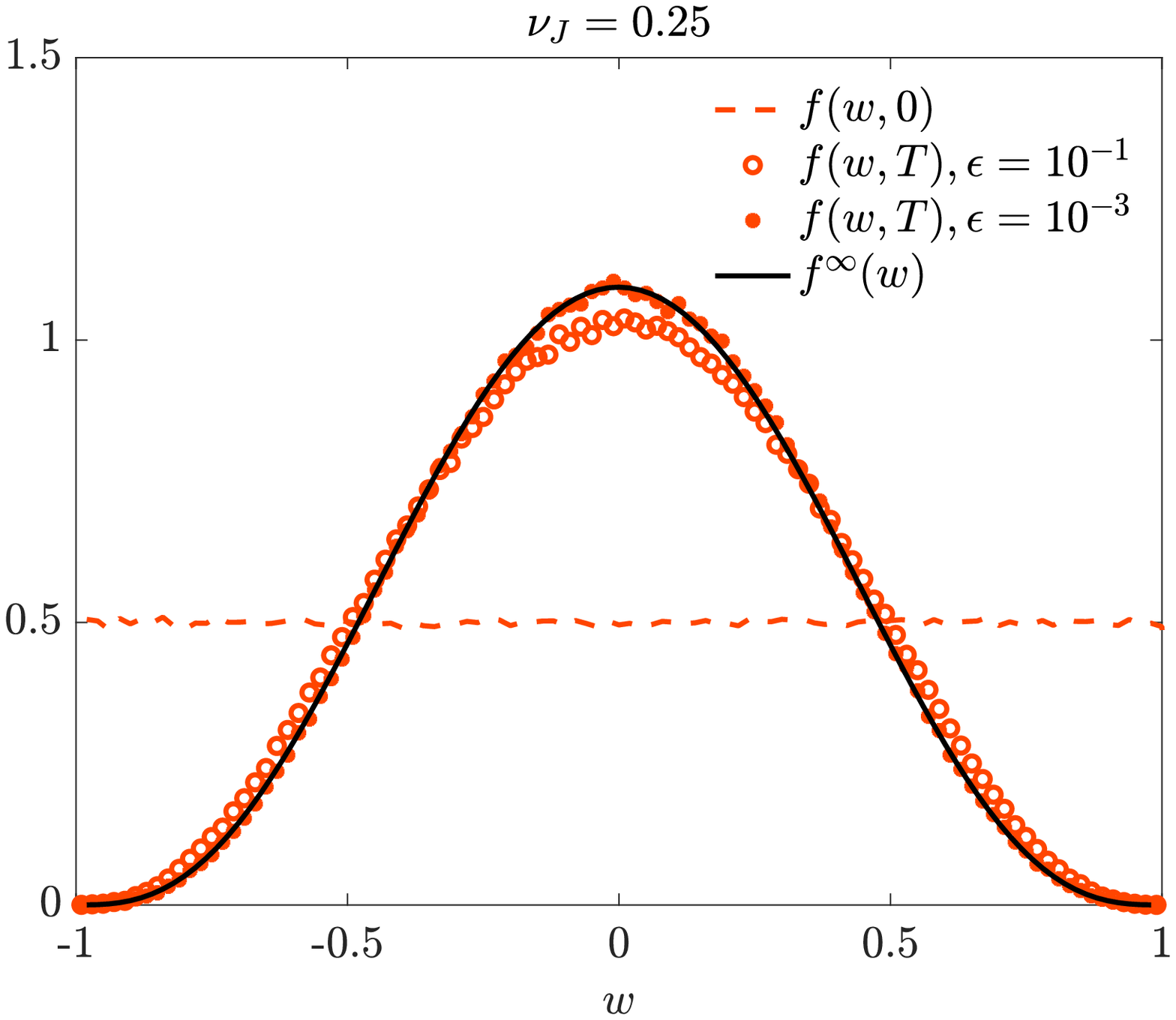}
\includegraphics[scale = 0.3]{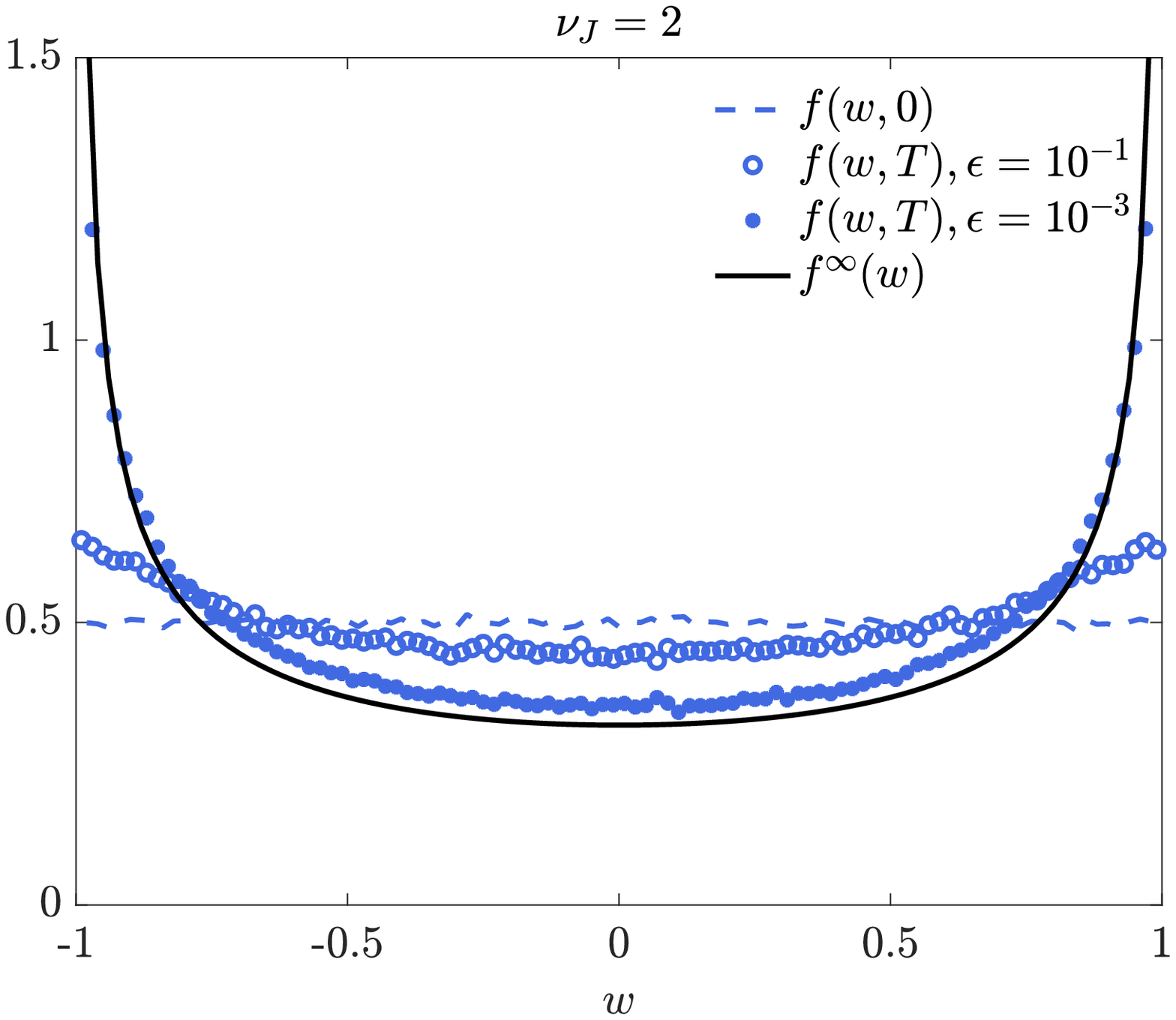} \\
\includegraphics[scale = 0.3]{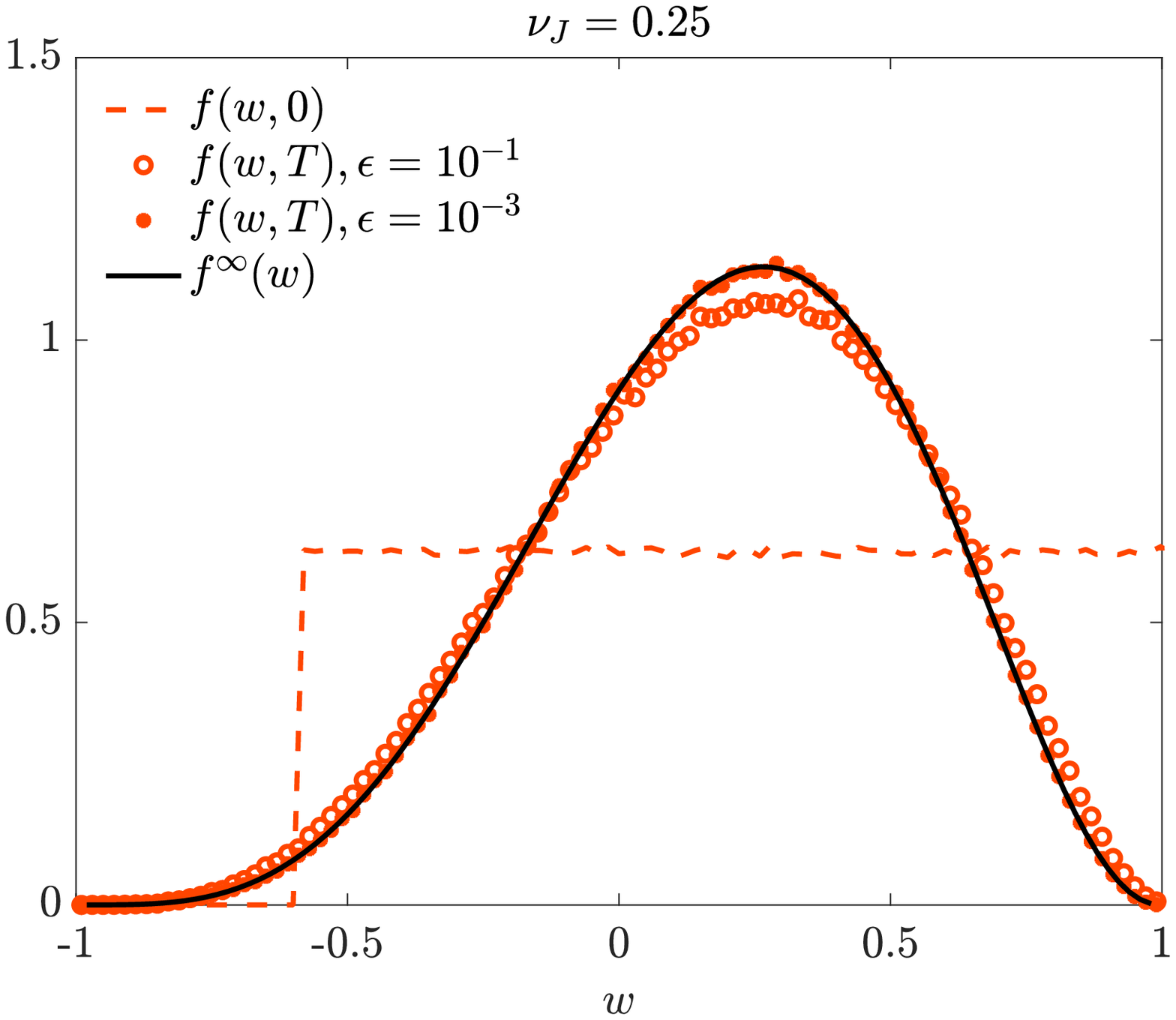}
\includegraphics[scale = 0.3]{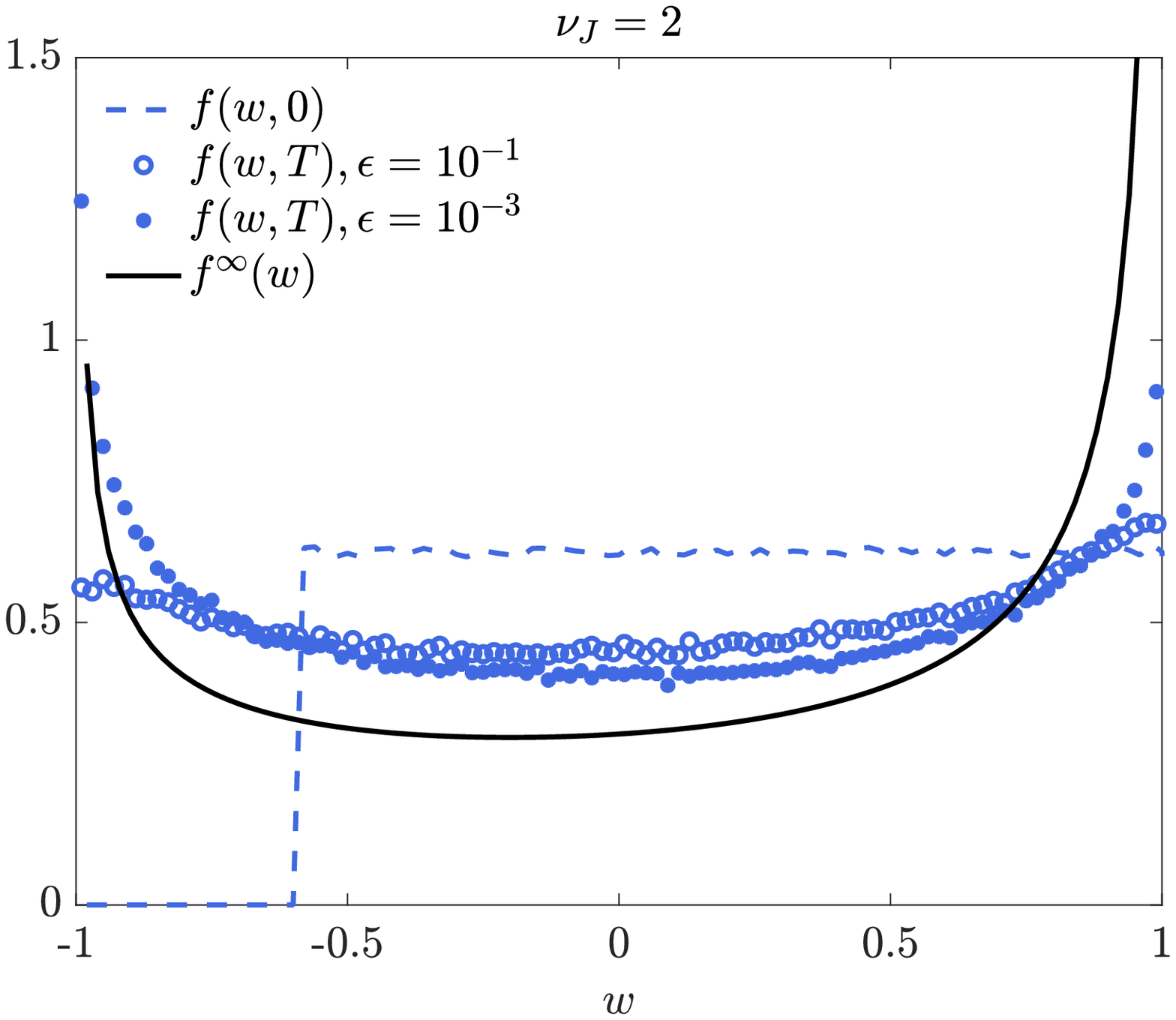}
\caption{\textbf{Test 1}. Comparison between DSMC solution of the Boltzmann-type problem \eqref{eq:kinetic_Qsum} and the Beta equilibrium solution of the Fokker-Planck model \eqref{eq:FP} for several values of $\nu_J = 0.25$ (left column) $\nu_J = 2$ (right column) and choices of the initial distribution. In particular we considered the choices in \eqref{eq:f01} (top row) and \eqref{eq:f02} (bottom row). The DSMC scheme has been implemented with $N = 10^6$ particles over the time frame $[0,5]$ with $\Delta t = \epsilon = 10^{-1}, 10^{-3}$.  } 
\label{fig:bol}
\end{figure}

In Figure \ref{fig:bol} we depict the densities reconstructed from the DSMC approach with $N = 10^6$ particles at time $T = 5$ and assuming $\Delta t = \epsilon = 10^{-3},10^{-1}$. In the top row we considered the initial distribution 
\begin{equation}
\label{eq:f01}
f_J(w,0) = 
\begin{cases}
\frac{1}{2} & w \in [-1,1] \\
0 & w \notin [-1,1]
\end{cases}
\end{equation}
such that $m_J(0) = \int_{-1}^1 f(w,0)dw = 0$ which is conserved in time. In the bottom row we consider the initial distribution 
\begin{equation}
\label{eq:f02}
f_J(w,0) = 
\begin{cases}
\frac{5}{8} & w \in [-0.6,1] \\
0 & w \notin [-0.6,1]
\end{cases}
\end{equation}
such that $m_J = 0.2$. We further assume that $\lambda_J = 1$ and $\sigma_J^2 = 0.25$ in the left column whereas $\sigma^2_J = 2$ in the right column. Hence, under the introduced choice of parameters we have considered $\nu_J = 0.25$ (left column) and $\nu_J = 2$ (right column). The emerging distribution is compared with the Beta distribution defined in  \eqref{eq:beta}. We may observe how, for decreasing values of $\epsilon\rightarrow 0^+$, we correctly approximate  the large time solution of the surrogate Fokker-Planck-type problem. 

\subsection{Test 2: consistency of the macroscopic limit}

In this test we compare the evolution of mass and local mean of the distributions $f_J$, $J\in \mathcal C$, solution to \eqref{eq:kinetic_opinion}, with the evolution of the obtained macroscopic system \eqref{eq:mass}-\eqref{eq:mean}. 

We are interested in the evolution $f_J(w,t)$, $J\in \mathcal C$, $w \in [-1,1]$, $t \ge 0$ solution to \eqref{eq:kinetic_opinion} and complemented by the initial condition $f_J(w,0) = f_J^0$. We consider a time discretization of the interval $[0,t_{\textrm{max}}]$ of size $\Delta t >0$. We denote by $f^n_J(w)$ the approximation of $f_J(w,t^n)$. Hence, we introduce a splitting strategy between the opinion consensus step $f^*_J = \mathcal O_{\Delta t}(f^n_J)$
\begin{equation}
\begin{cases}
\partial_t f_J^* = \dfrac{1}{\tau} \bar Q_J(f_J^*,f_J^*), \\
f_J^*(w,0) = f_J^n(w), \qquad J \in \mathcal C
\end{cases}
\label{eq:split_O}
\end{equation}
and the epidemiological step $f^{**}_J = \mathcal E_{\Delta t}(f^{**}_J)$
\begin{equation}
\begin{cases}
\partial_t f_S^{**} = -f_S^{**}(1-w)\rho_I^{**}(1-m_I^{**}) \\
\partial_t f_E^{**} = f_S^{**}(1-w)\rho_I^{**}(1-m_I^{**}) - \sigma_E f_E^{**} \\
\partial_t  f_I^{**} = \sigma_E f_E^{**} - \gamma f_I^{**} \\
\partial_t f_R^{**} = \gamma f_I^{**}, \\
f^{**}_J(w,0) = f^*_J(w,\Delta t). 
\end{cases}
\label{eq:split_E}
\end{equation}
The operator $\bar Q_J(\cdot,\cdot)$ in \eqref{eq:split_O} has been defined in \eqref{eq:FP} together with no-flux boundary conditions. Hence, the solution at time $t^{n+1}$ is given by the combination of the two described steps. In particular a first order splitting strategy corresponds to 
\[
f_J^{n+1}(w) = \mathcal E_{\Delta t}(\mathcal O_{\Delta t}(f_J^n(w))),
\]
whereas the second order Strang splitting method is obtained as
\[
f_J^{n+1}(w) = \mathcal E_{\Delta t/2}(\mathcal O_{\Delta t}(\mathcal E_{\Delta t/2}(f_J^n(w)))),
\]
for all $J \in \mathcal C$.
The opinion consensus step \eqref{eq:split_O} is solved by means of a second-order semi-implicit structure-preserving (SP) method for Fokker-Planck equations, see \cite{PZ}. The integration of the epidemiological  step \eqref{eq:split_E} is performed with an RK4 method. In the following, we will adopt a Strang splitting approach.

We consider the following artificial parameters characterizing the epidemiological dynamics $\beta = 0.4$, $\sigma_E = 1/2$, $\gamma = 1/12$. These values are strongly dependent on the infectious disease under investigation. We highlight that, without having the intention to use real data for the calibration of the presented model, these values are coherent with several recent works for the COVID-19 pandemic \cite{ABBDPTZ,BDM,DTZ}.

\subsection{Test 2a: equilibrium closure}
In this test we assume a constant interaction function $P(\cdot,\cdot) \equiv 1$ such that the Fokker-Planck model is characterized by a Beta equilibrium distribution \eqref{eq:beta} as shown in Section \ref{subsect:FP}.
To define the initial condition we introduce the distributions 
\[
g(w) =
\begin{cases}
1 & w \in [-1,0] \\
0 & \textrm{elsewhere},
\end{cases}
\qquad
h(w)=
\begin{cases}
1 & w \in [0,1] \\
0 & \textrm{elsewhere},
\end{cases}
\]
and we consider 
\begin{equation}
\label{eq:init}
\begin{split}
f_S(w,0) = \rho_S(0)g(w),\qquad f_E(w,0) = \rho_E(0)g(w),\\
f_I(w,0) = \rho_I(0)h(w),\qquad f_R(w,0) = \rho_R(0)h(w),
\end{split}
\end{equation}
with $\rho_E(0) =\rho_I(0) = \rho_R(0) = 10^{-2}$ and $\rho_S = 1-\rho_E(0)-\rho_I(0)-\rho_R(0)$.

\begin{figure}
\centering
\includegraphics[scale = 0.3]{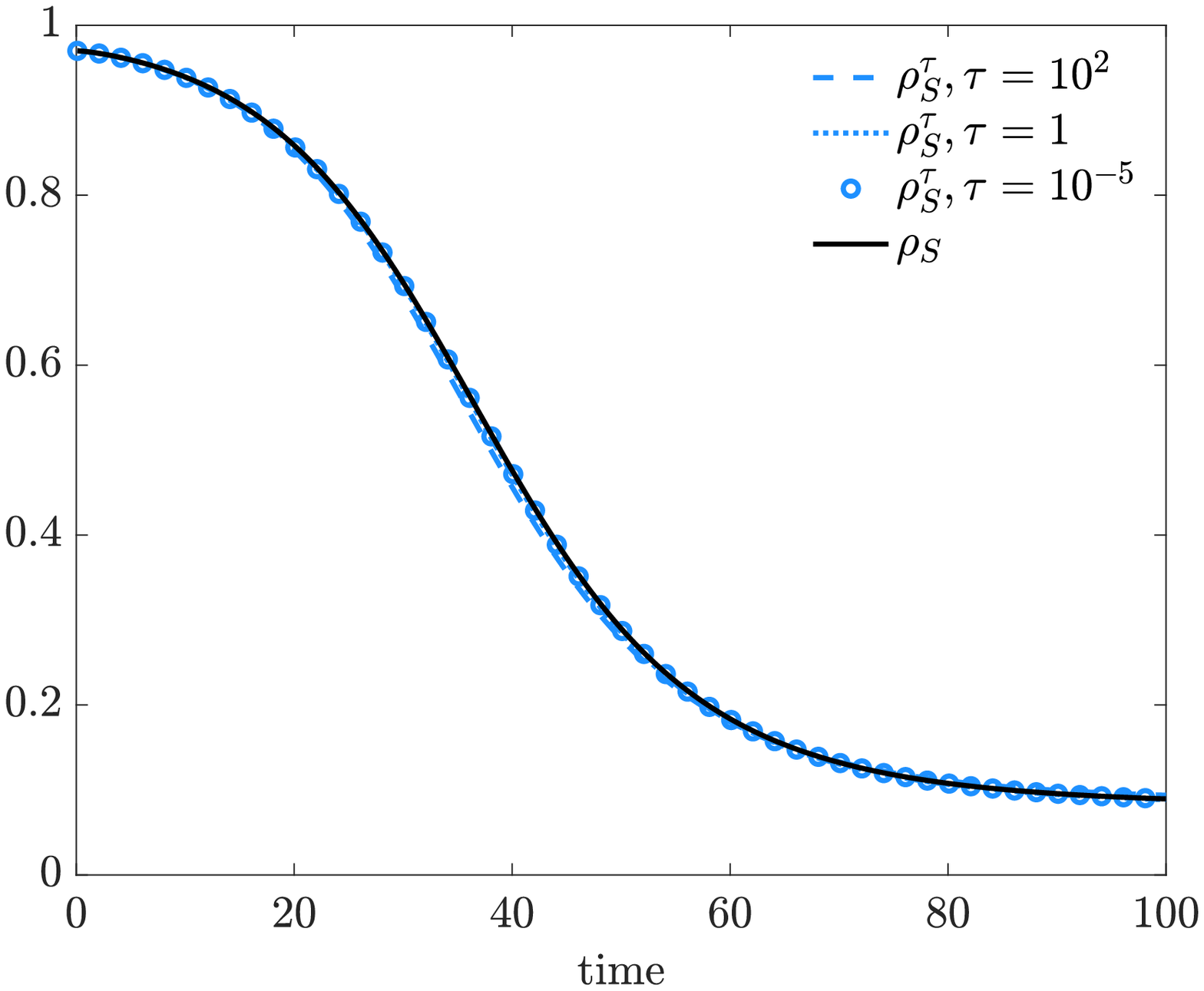}
\includegraphics[scale = 0.3]{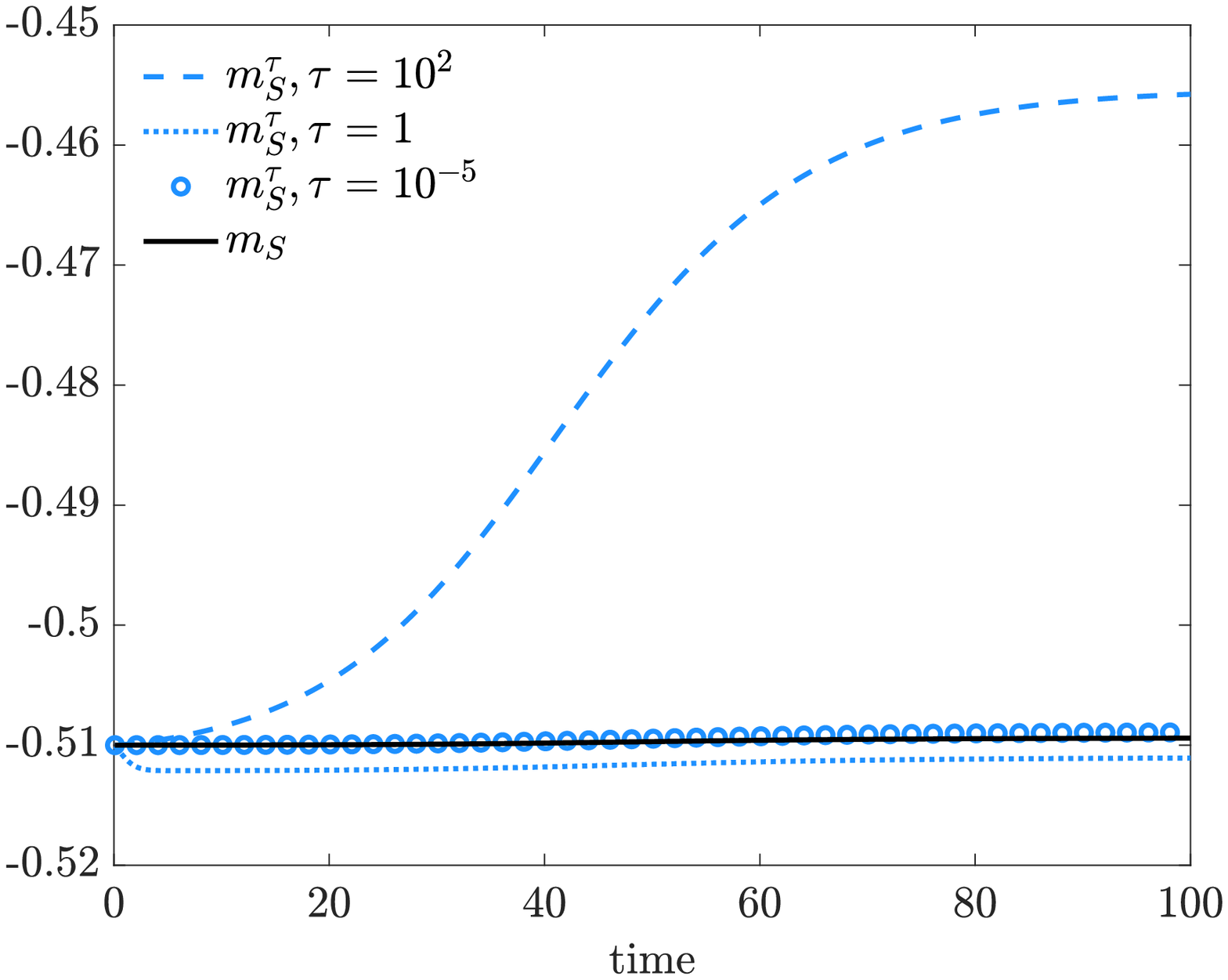} \\
\includegraphics[scale = 0.3]{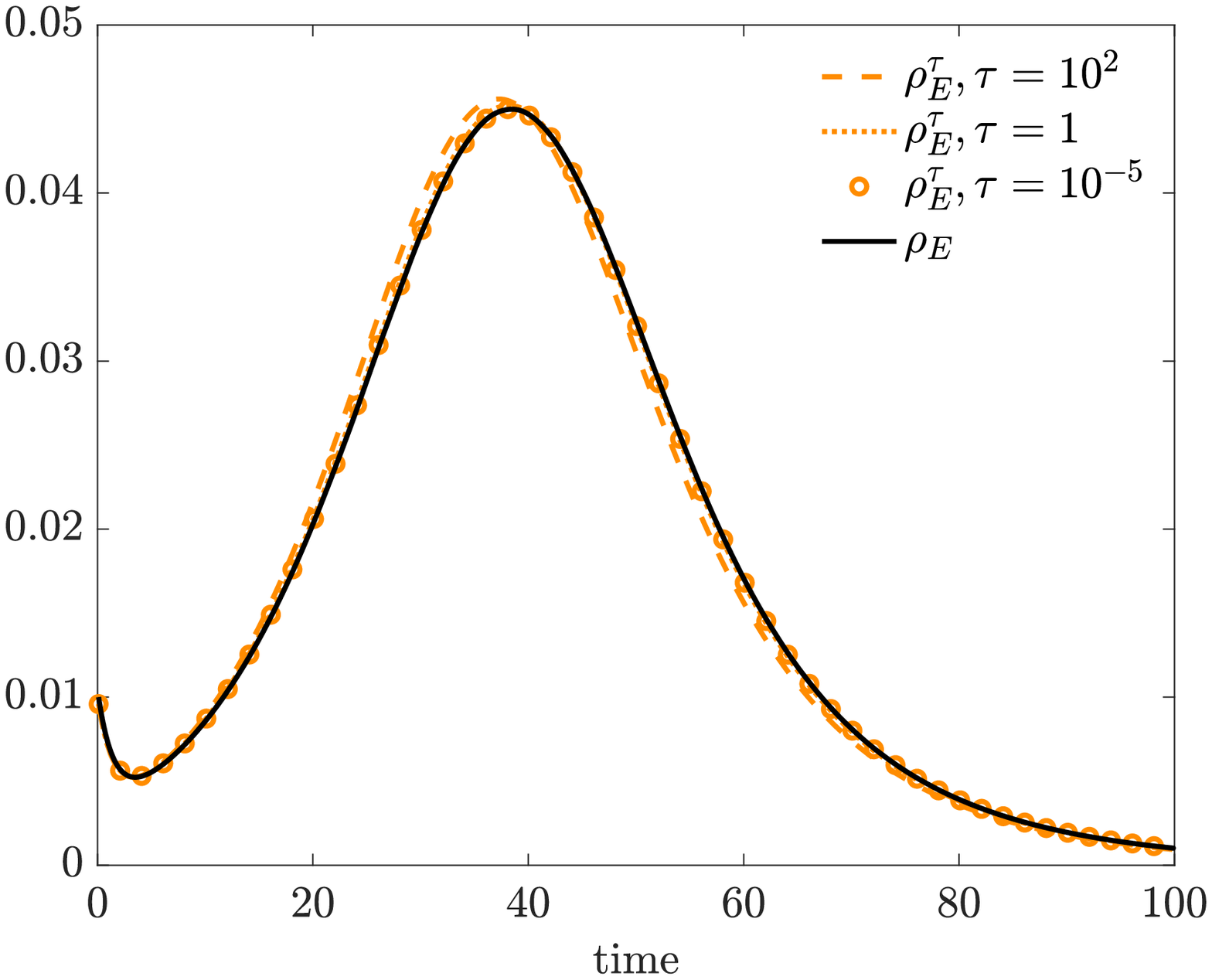}
\includegraphics[scale = 0.3]{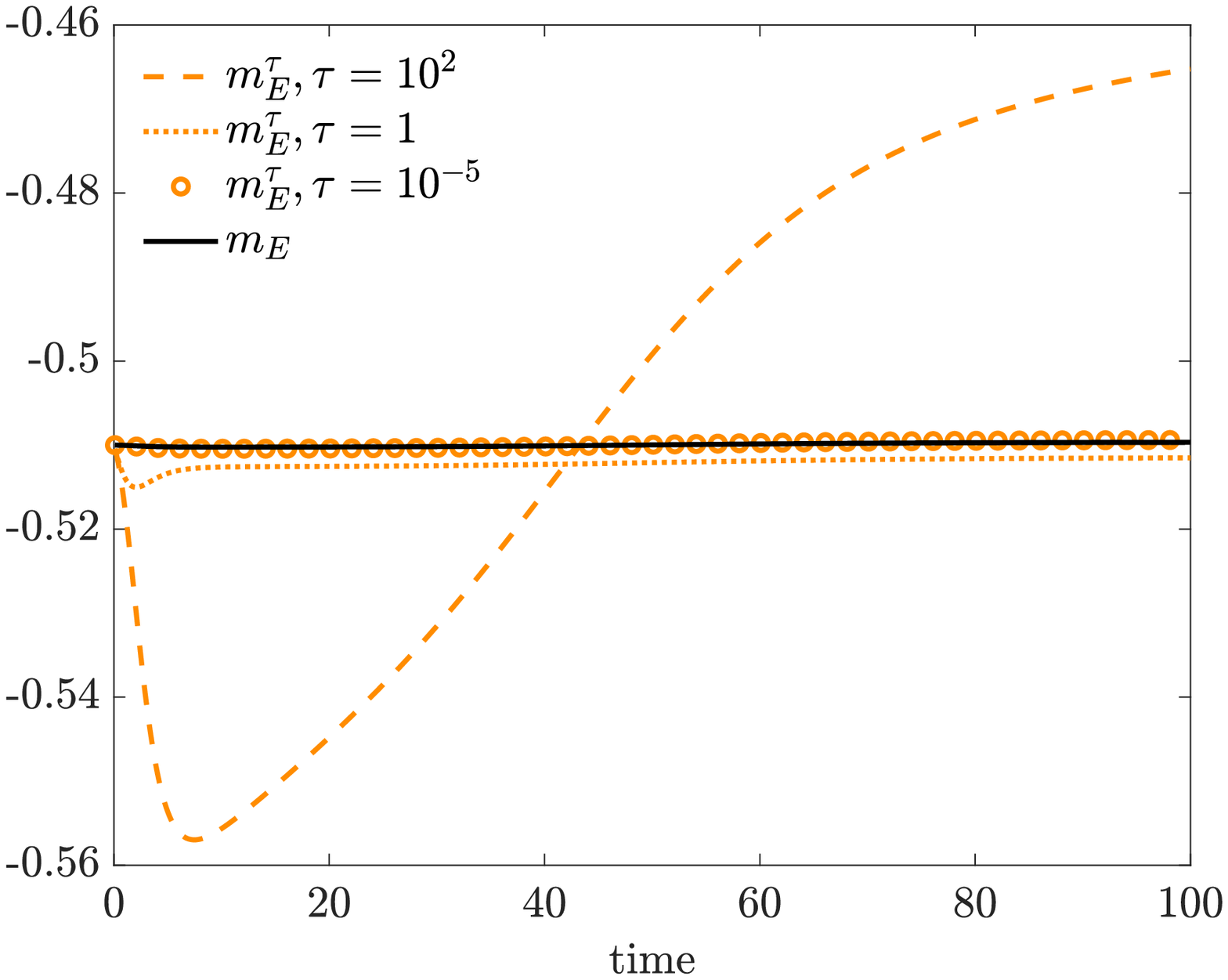} \\
\includegraphics[scale = 0.3]{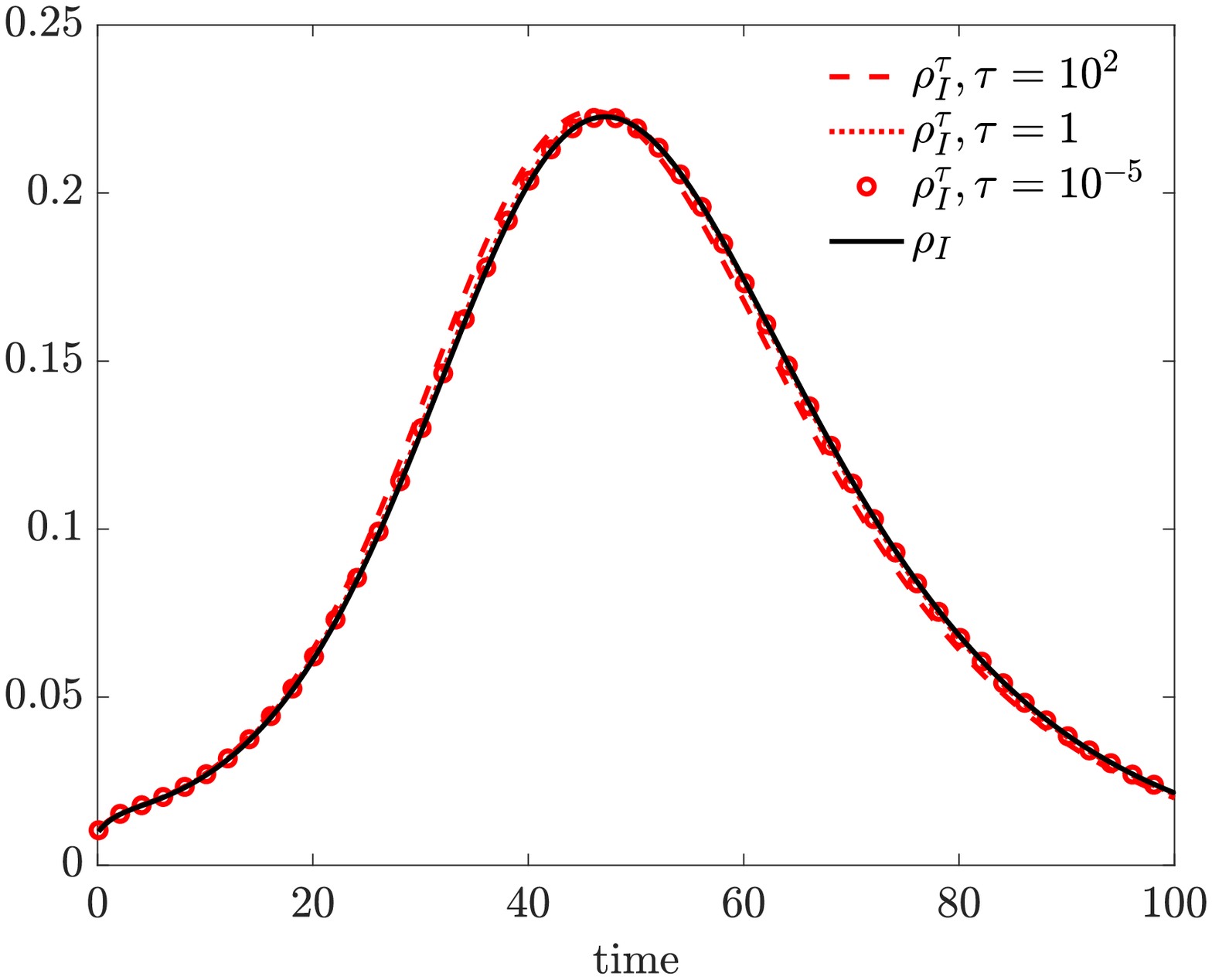}
\includegraphics[scale = 0.3]{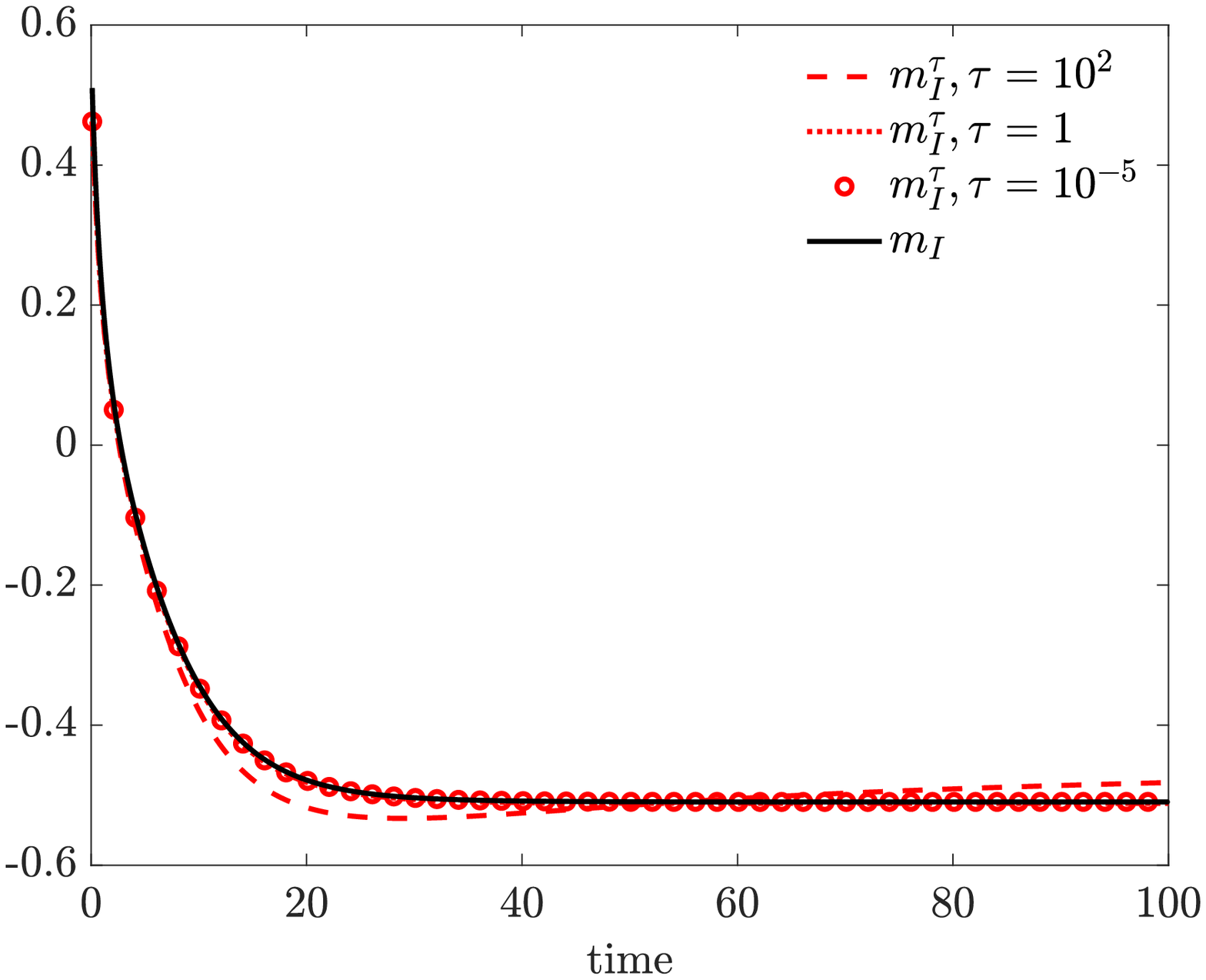}\\
\includegraphics[scale = 0.3]{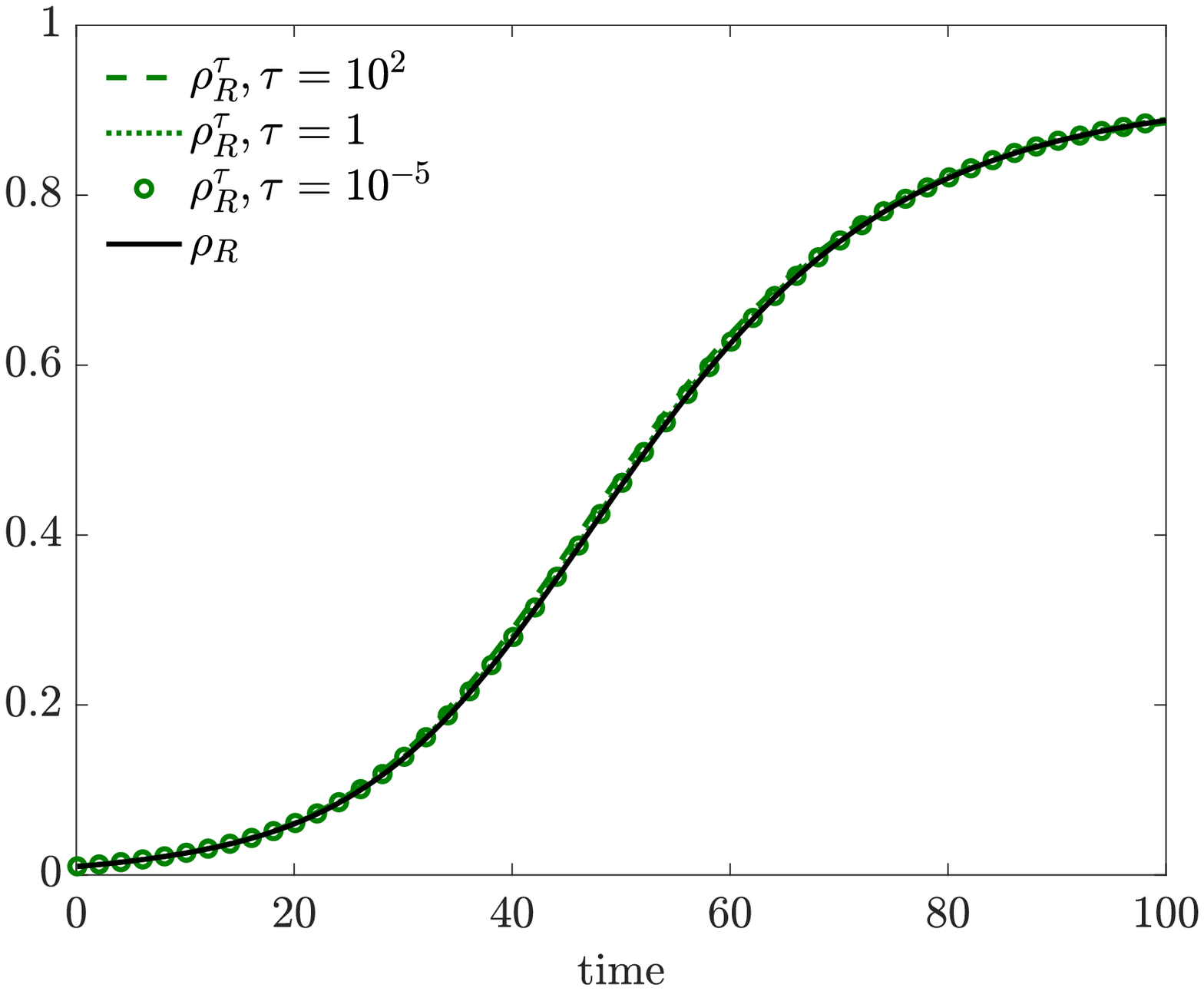}
\includegraphics[scale = 0.3]{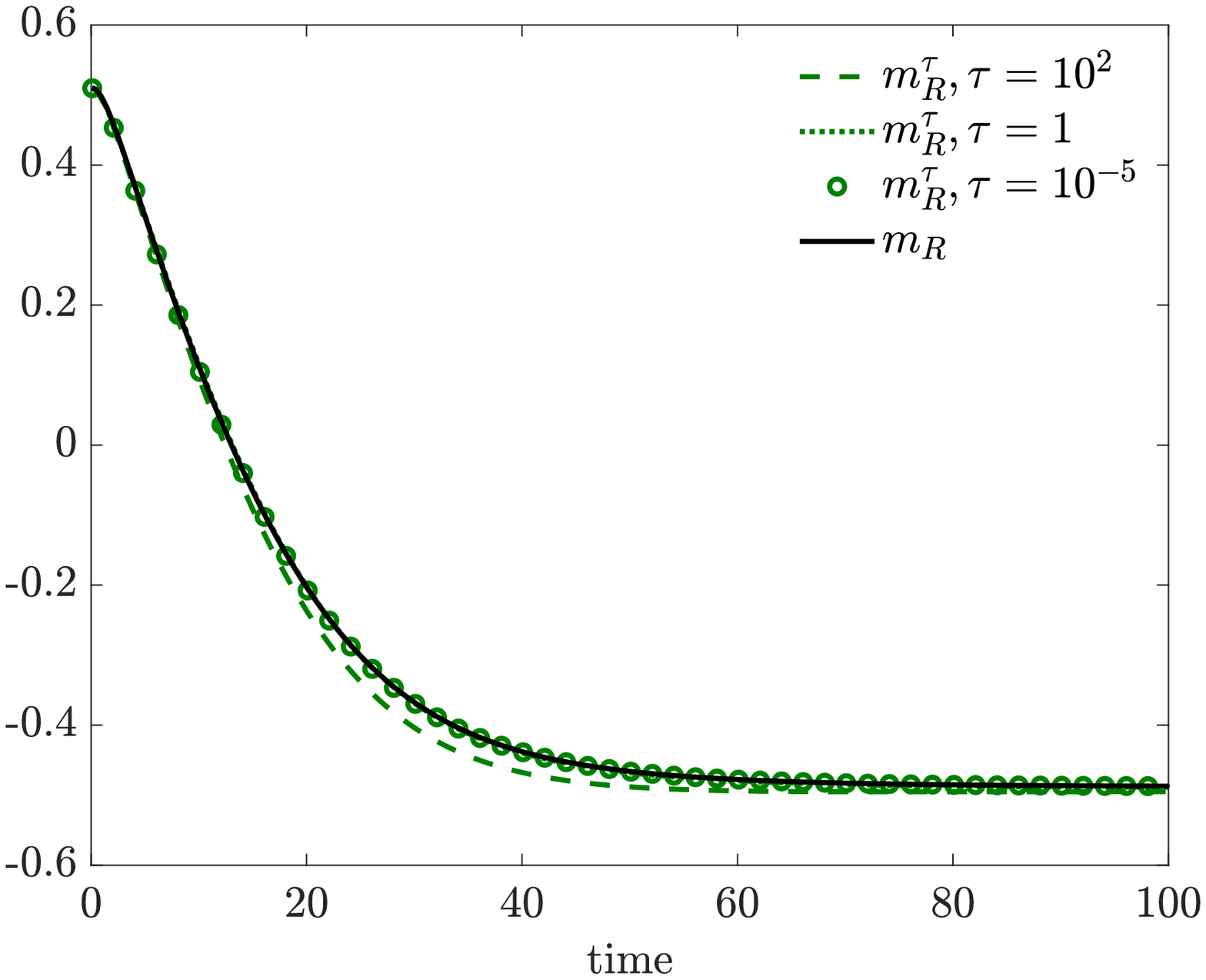}
\caption{\textbf{Test 2a}. Evolution of the macroscopic quantities defined in  \eqref{eq:mass}-\eqref{eq:mean} and the ones extrapolated from the kinetic model \eqref{eq:kinetic_opinion} for several values $\tau = 10^{-5},1,10^2$, see \eqref{eq:macro_tau}. Discretization of the domain $[-1,1]$ obtained with $N_w = 201$ gridpoints, discretization of the time frame $[0,100]$ obtained with $\Delta t = 10^{-1}$. The initial distributions have been defined in  \eqref{eq:init} whereas we fixed $\lambda_J = 1$ and $\sigma^2_J = 10^{-3}$ for all $J \in \mathcal C$.   }
\label{fig:limit}
\end{figure}

We solve numerically \eqref{eq:split_O}-\eqref{eq:split_E} over the time frame $[0,t_{\textrm{max}}]$ and we introduce the grid $w_i \in [-1,1]$ with $w_{i+1}-w_i$, where $\Delta w>0$, $i= 1,\dots,N_w$. We introduce also a time discretization such that $t^{n} = n\Delta t$, $\Delta t >0$, and $n = 0,\dots,T$ with $T\Delta t = t_{\textrm{max}}$. For all the details on the considered numerical scheme we point the interested reader to \cite{PZ}. Hence, for several values of $\tau>0$, we compare the evolution of the computed observable quantities defined as
\begin{equation}
\label{eq:macro_tau}
\rho^{\tau}_J(t) = \int_{-1}^1 f_J(w,t)dw, \qquad m_J^{\tau}(t) = \dfrac{1}{\rho_J^{\tau}(t)} \int_{-1}^1 w f_J(w,t)dw
\end{equation}
with the ones in \eqref{eq:mass}-\eqref{eq:mean} whose dynamics has been determined through a suitable kinetic closure in the limit $\tau\to 0^+$. In \eqref{eq:macro_tau} we have highlighted the dependence on the scale parameter $\tau>0$ through a superscript. It is important to remark that the introduced closure strategy is essentially based on the assumption that opinion dynamics are faster than the ones characterizing the epidemic. Furthermore, we fix as initial values of the coupled system \eqref{eq:mass}-\eqref{eq:mean} the values  $\rho_J(0)$ and $m_J(0) $, for all $J\in \mathcal C$. 

In Figure \ref{fig:limit} we present the evolution of the macroscopic system \eqref{eq:mass}-\eqref{eq:mean} and of the observable quantities \eqref{eq:macro_tau} for several $\tau = 10^{-5},1,100$. The  consensus dynamics is characterized by $\lambda_J = 1$, $\sigma_J^2 = 10^{-3}$ for all $J \in \mathcal C$, such that $\nu_S = 10^{-3}$. We can easily observe how, for small values of $\tau \ll 1$, the macroscopic model obtained through a Beta-type equilibrium closure is coherent with the evolution of mass and mean of the kinetic model \eqref{eq:kinetic_opinion}. 

\begin{figure}
\centering
\includegraphics[scale = 0.3]{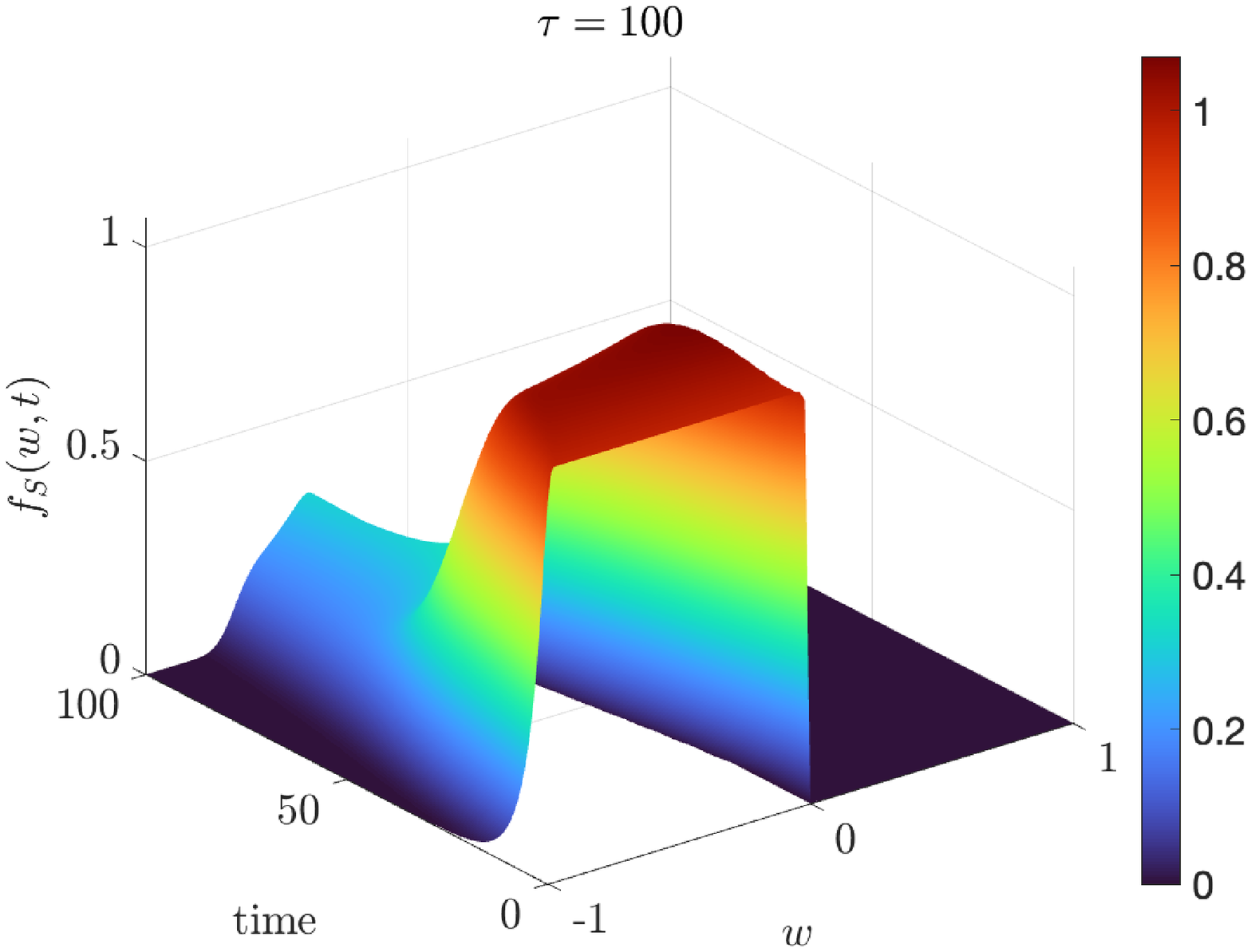}
\includegraphics[scale = 0.3]{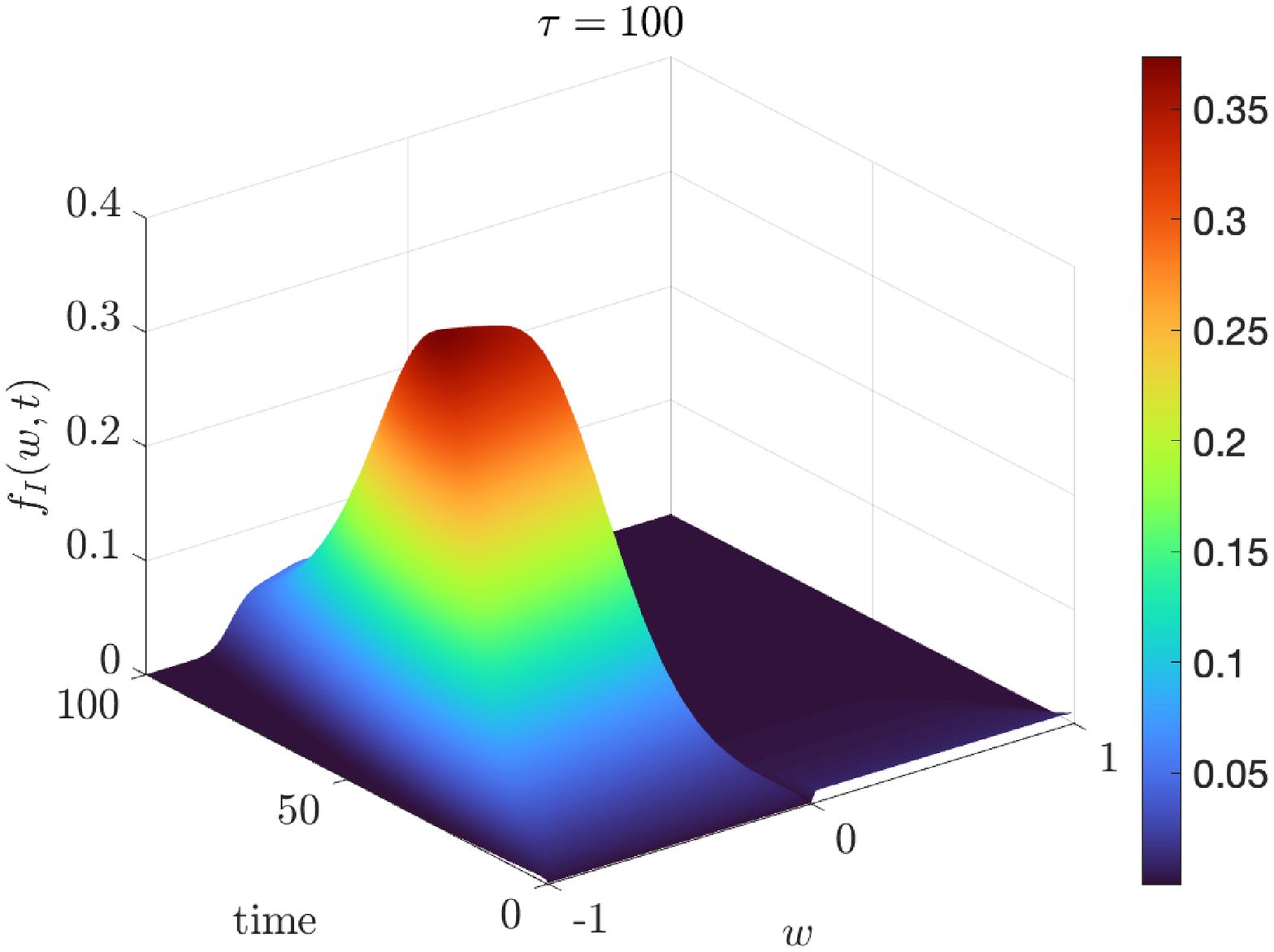} \\
\includegraphics[scale = 0.3]{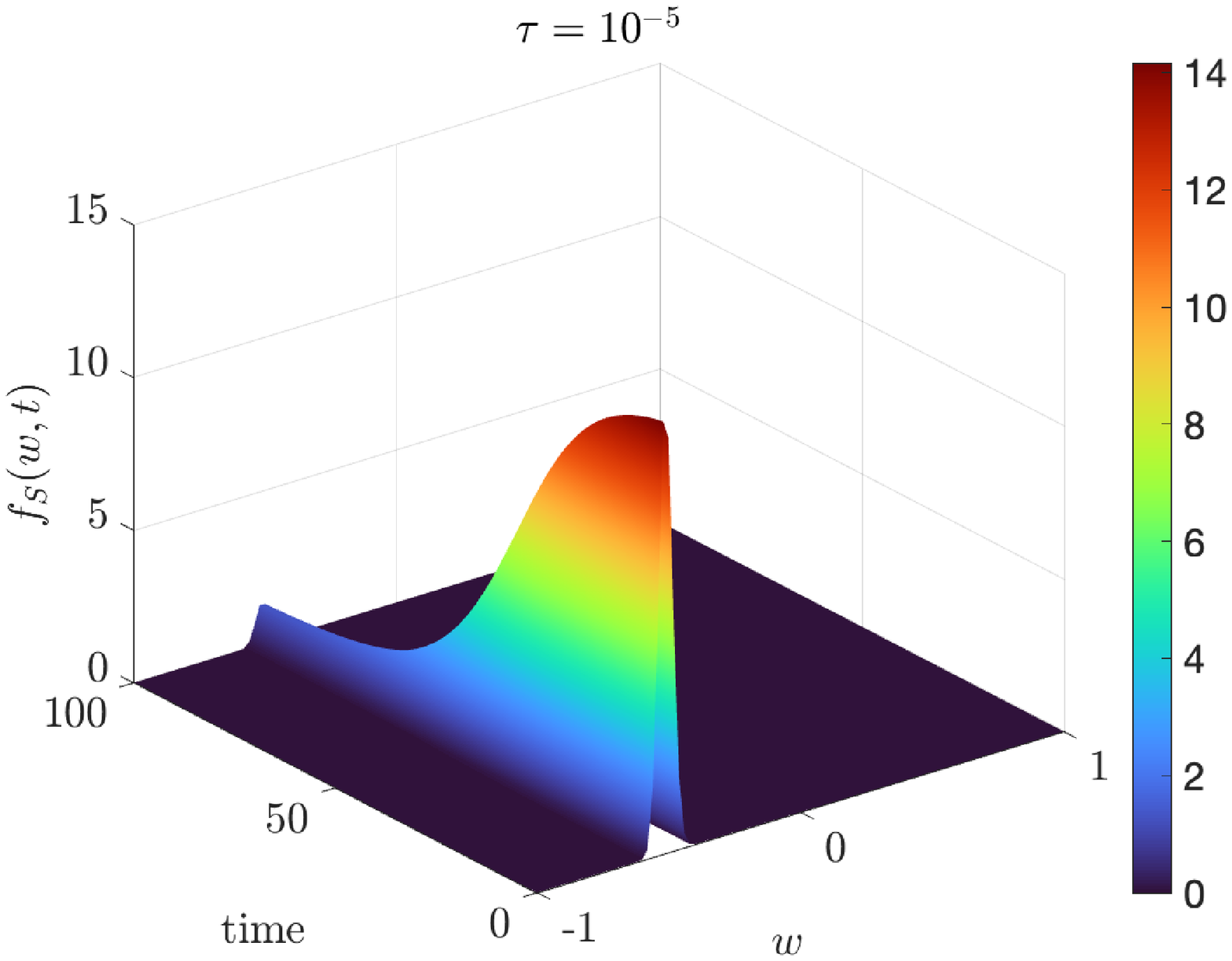}
\includegraphics[scale = 0.3]{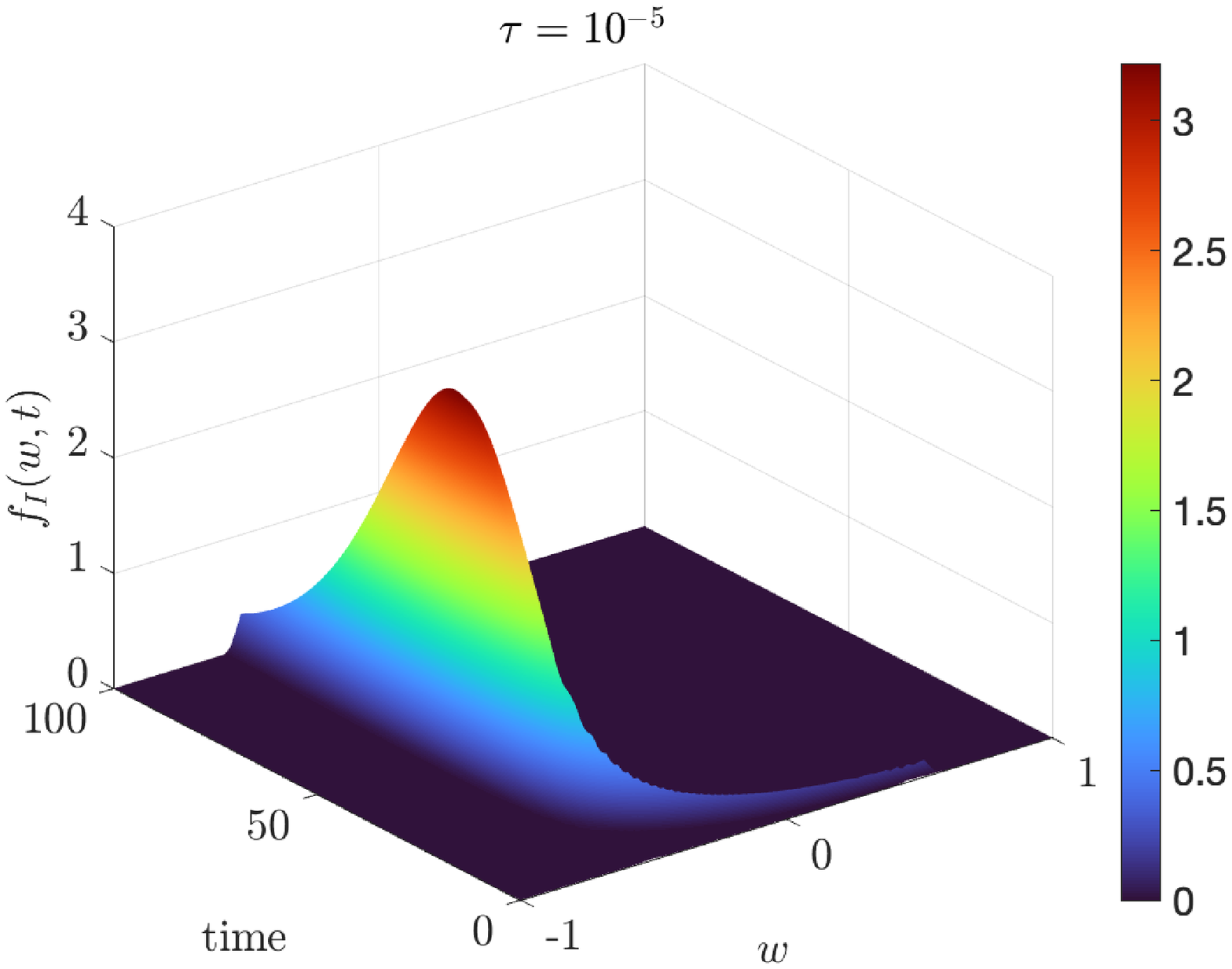}
\caption{\textbf{Test 2a.} Evolution of the kinetic distributions $f_S$ and $f_I$ over the time interval $[0,100]$ for $\tau = 100$ (top row) and $\tau = 10^{-5}$ (bottom row). The epidemic dynamics have been characterized by $\beta = 0.4$,$\sigma_E = 1/2$, $\gamma = 1/12$. The solution of the Fokker-Planck step \eqref{eq:split_O} has been performed through a semi-implicit SP scheme over the a grid of $N_w = 201$ nodes and $\Delta t = 10^{-1}$. Initial distributions defined in  \eqref{eq:init}.}
\label{fig:limit_dist}
\end{figure}

In Figure \ref{fig:limit_dist} we show the evolution of the kinetic distributions $f_S(w,t)$ and $f_I(w,t)$ for $t \in [0,100]$. The parameters characterizing the opinion and epidemic dynamics are coherent with the ones chosen for Figure \ref{fig:limit}. We may easily observe how for $\tau = 100$ the distributions are far from the Beta equilibrium \eqref{eq:beta} whereas for $\tau = 10^{-5}$ the kinetic distributions $f_J$ are of Beta-type. Therefore, for small $\tau\ll1$, the opinion exchanges are faster than the epidemic dynamics and we are allowed to assume a Beta-type closure  as in \eqref{eq:m2}.

\subsection{Test2b: the bounded confidence case}\label{subsect:test2b}

In this test we consider an interaction function of the form 
\begin{equation}
\label{eq:BC}
P(w,w_*) = \chi(|w-w_*|\le \Delta),\qquad w,w_* \in [-1,1],
\end{equation}
where $\chi(\cdot)$ is the indicator function, and $\Delta  \in [0,2]$ is a confidence threshold parameter above which the agents' with opinions $w$ and $w_*$ do not interact. In the case $\Delta = 0$ only agents sharing the same opinion interact, whereas for $\Delta = 2$ the interaction function is such that $P(\cdot,\cdot)\equiv 1$ since $|w-w_*|\le 2$ for all $w,w_* \in [-1,1]$. Bounded confidence-type dynamics  have been introduced in \cite{HK} and have been studied to observe the loss of global consensus. Indeed, for large times, the agents' opinion form several clusters whose number and size depends on the parameter $\Delta >0$ and the initial opinions. We highlight that, since bounded confidence interactions \eqref{eq:BC} are symmetric, the mean opinion is preserved in time \cite{PTTZ}.

\begin{figure}
\centering
\includegraphics[scale = 0.3]{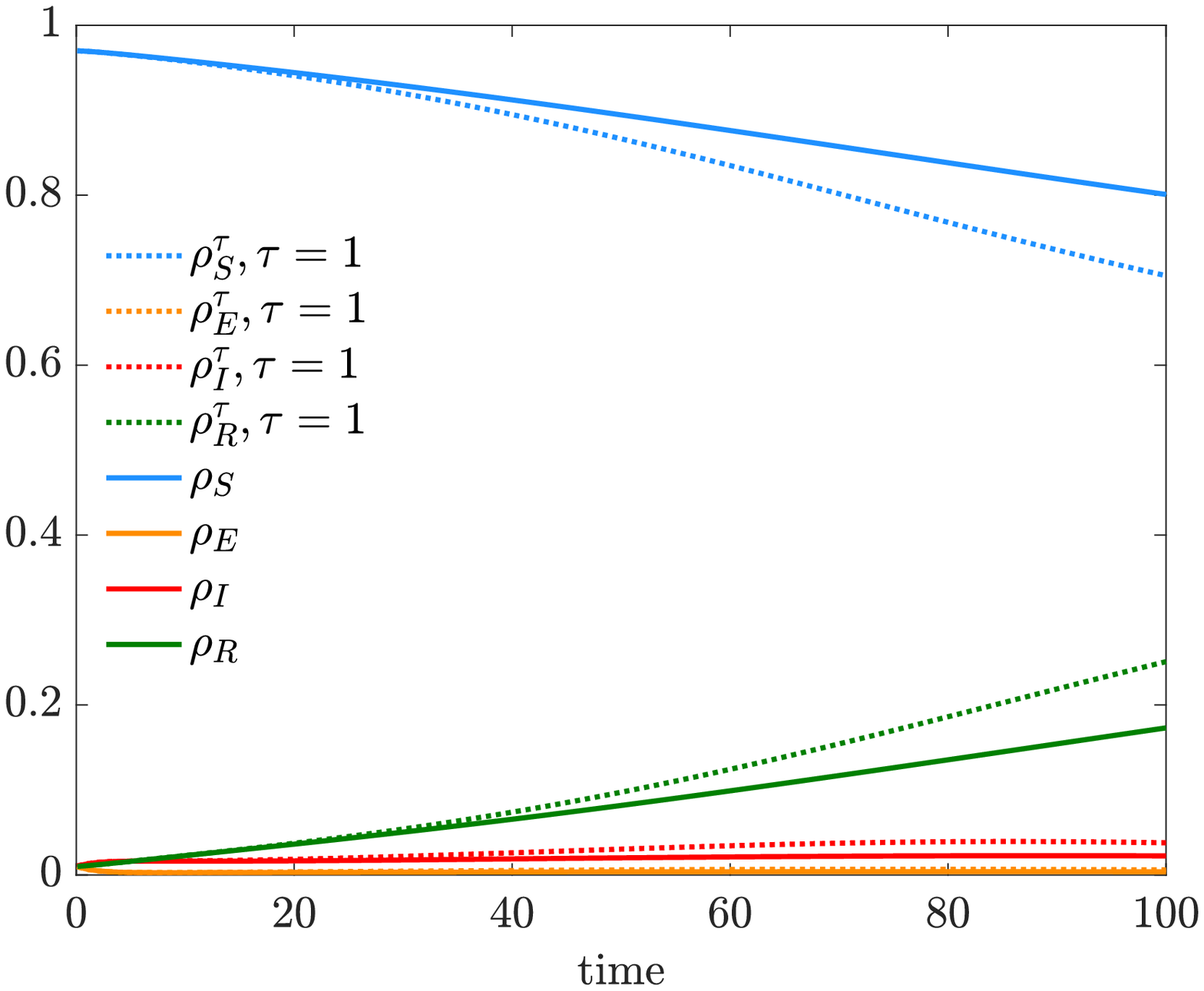} 
\includegraphics[scale = 0.3]{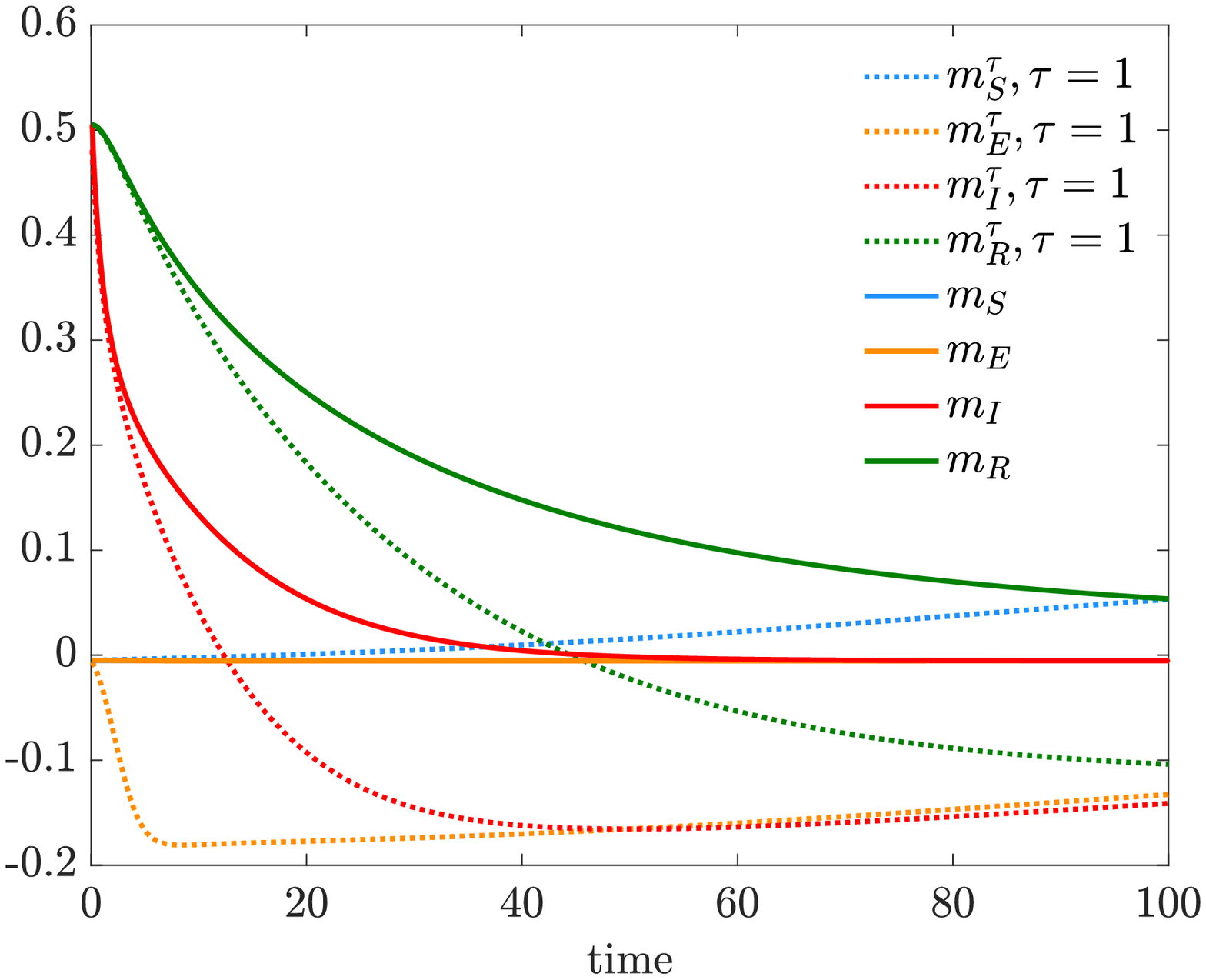}\\
\includegraphics[scale = 0.3]{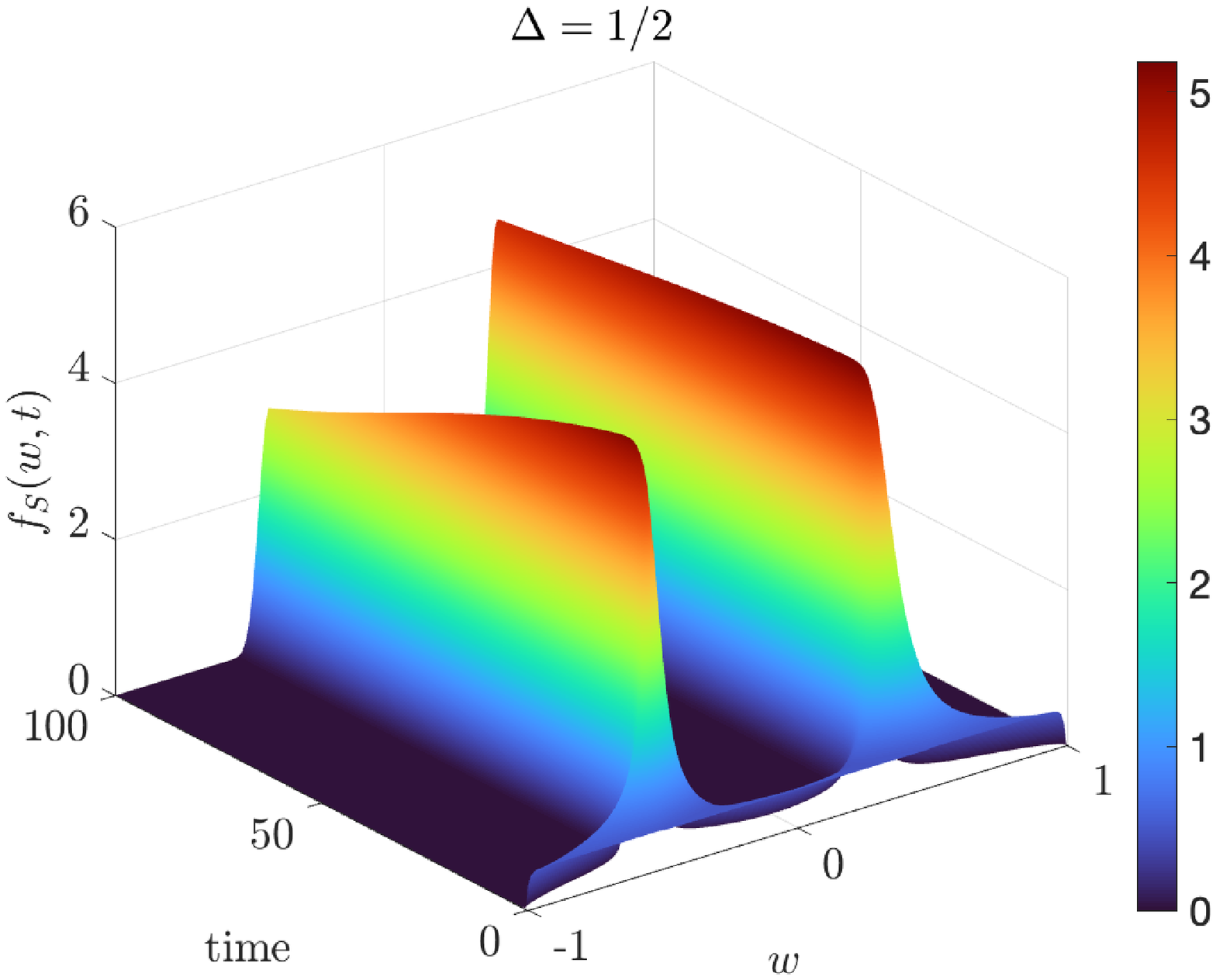}
\includegraphics[scale = 0.3]{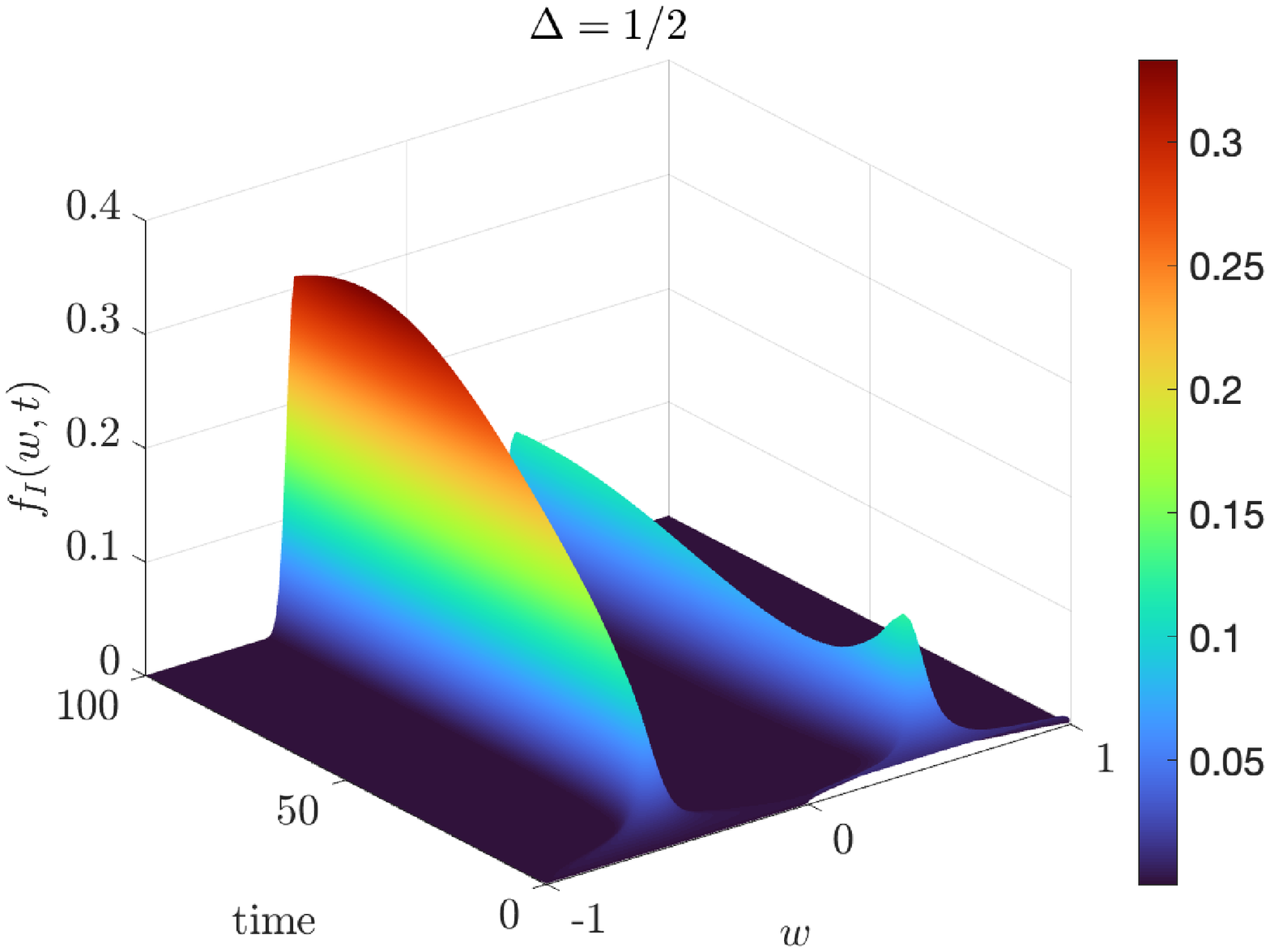}
\caption{\textbf{Test 2b}. We consider a the bounded confidence interaction function \eqref{eq:BC} with $\Delta = \frac{1}{2}$. Top row: evolution of mass fractions (left) and mean values (right) for the agents in compartments $\mathcal C$ with $\tau = 1$ and extrapolated from the kinetic model \eqref{eq:kinetic_opinion} with a Fokker-Planck operator $\bar Q(\cdot,\cdot)(w,t)$ of the form \eqref{eq:Q_BC}. Bottom row: evolution of the kinetic distributions for the compartments $S,I \in \mathcal C$. The solution of the Fokker-Planck step \eqref{eq:split_O} has been performed through a semi-implicit SP scheme over the a grid of $N_w = 201$ gridpoints and $\Delta t = 10^{-1}$. Initial distributions defined in  \eqref{eq:init2}.}
\label{fig:BC12}
\end{figure}

\begin{figure}
\centering
\includegraphics[scale = 0.3]{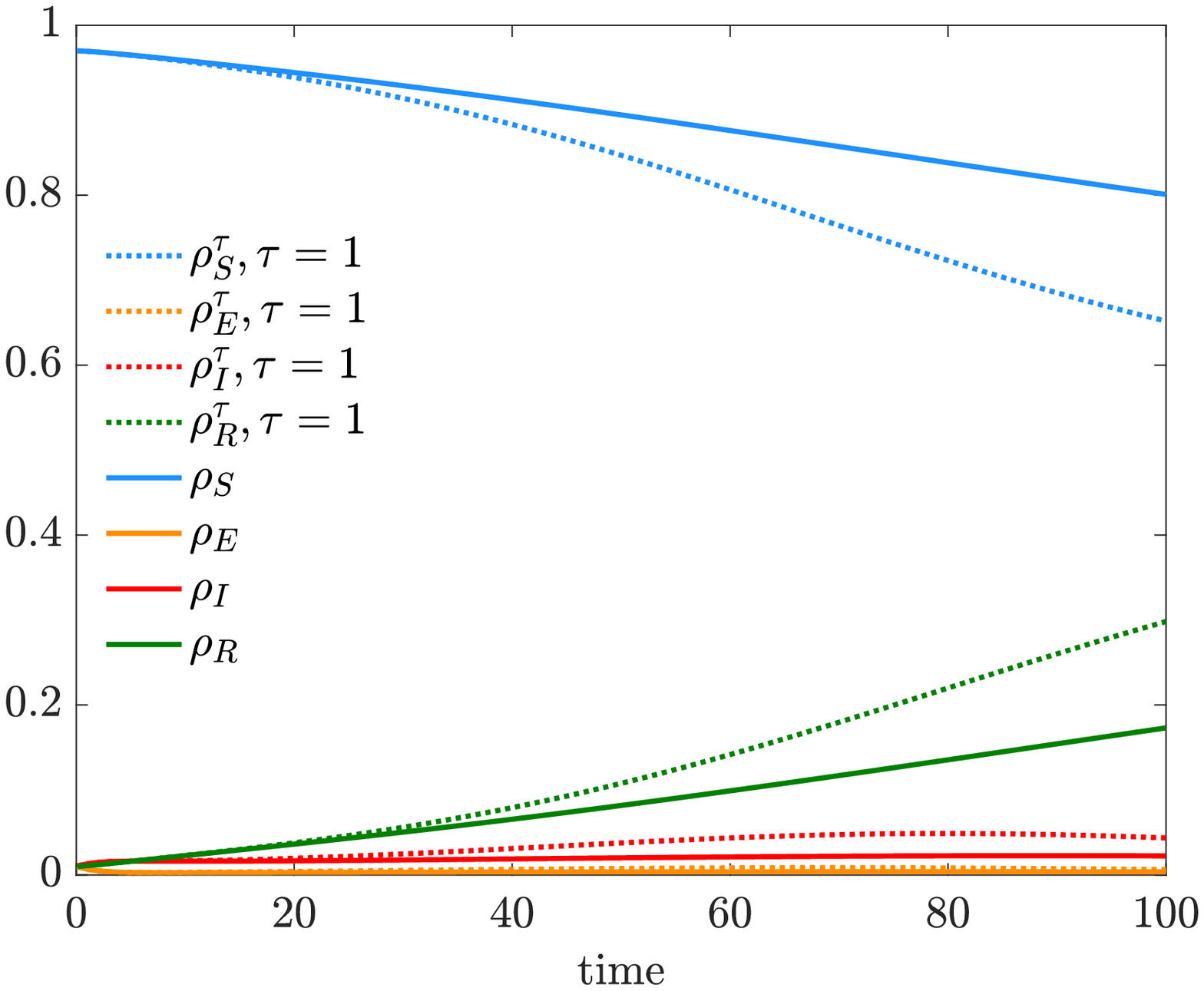} 
\includegraphics[scale = 0.3]{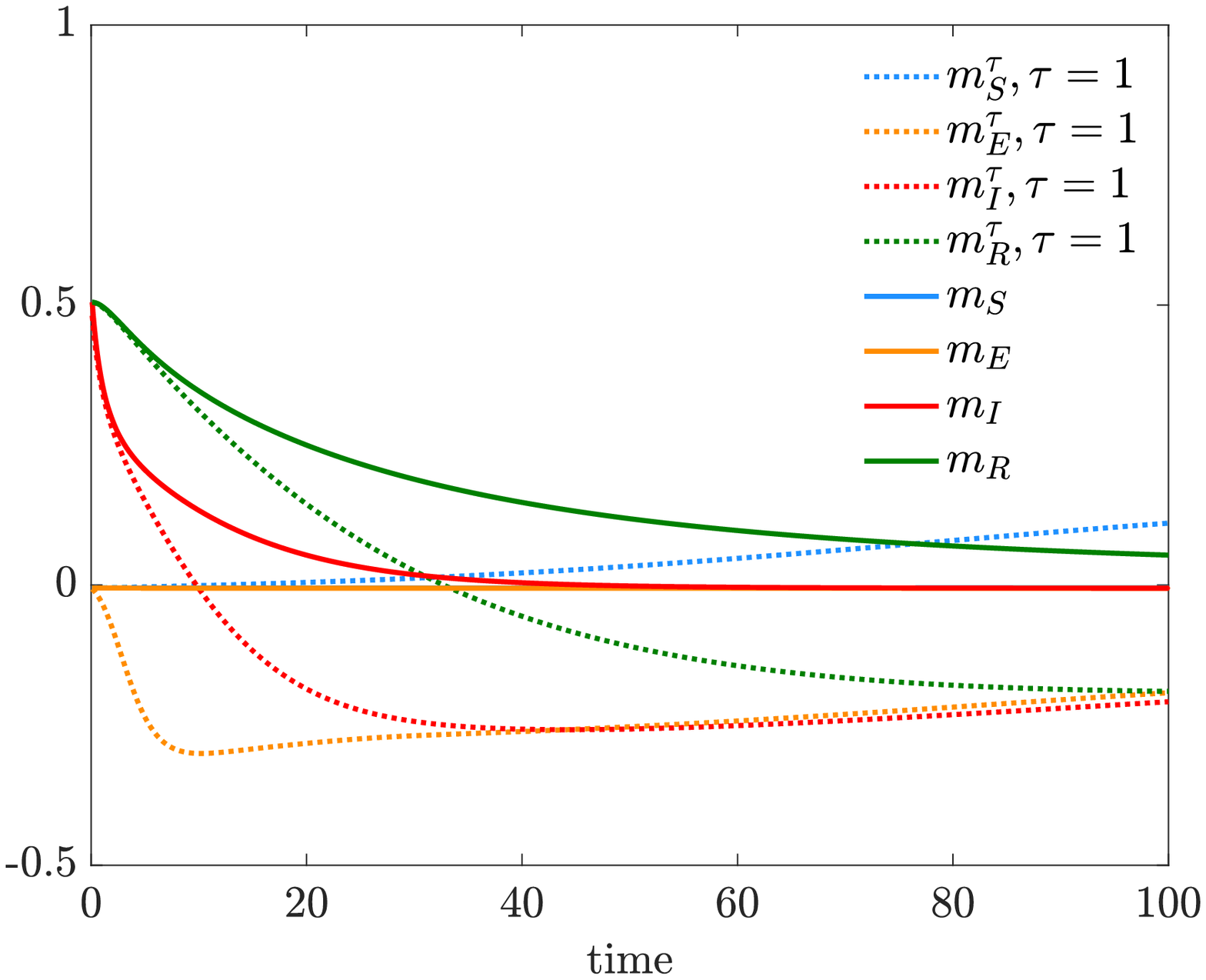}\\
\includegraphics[scale = 0.3]{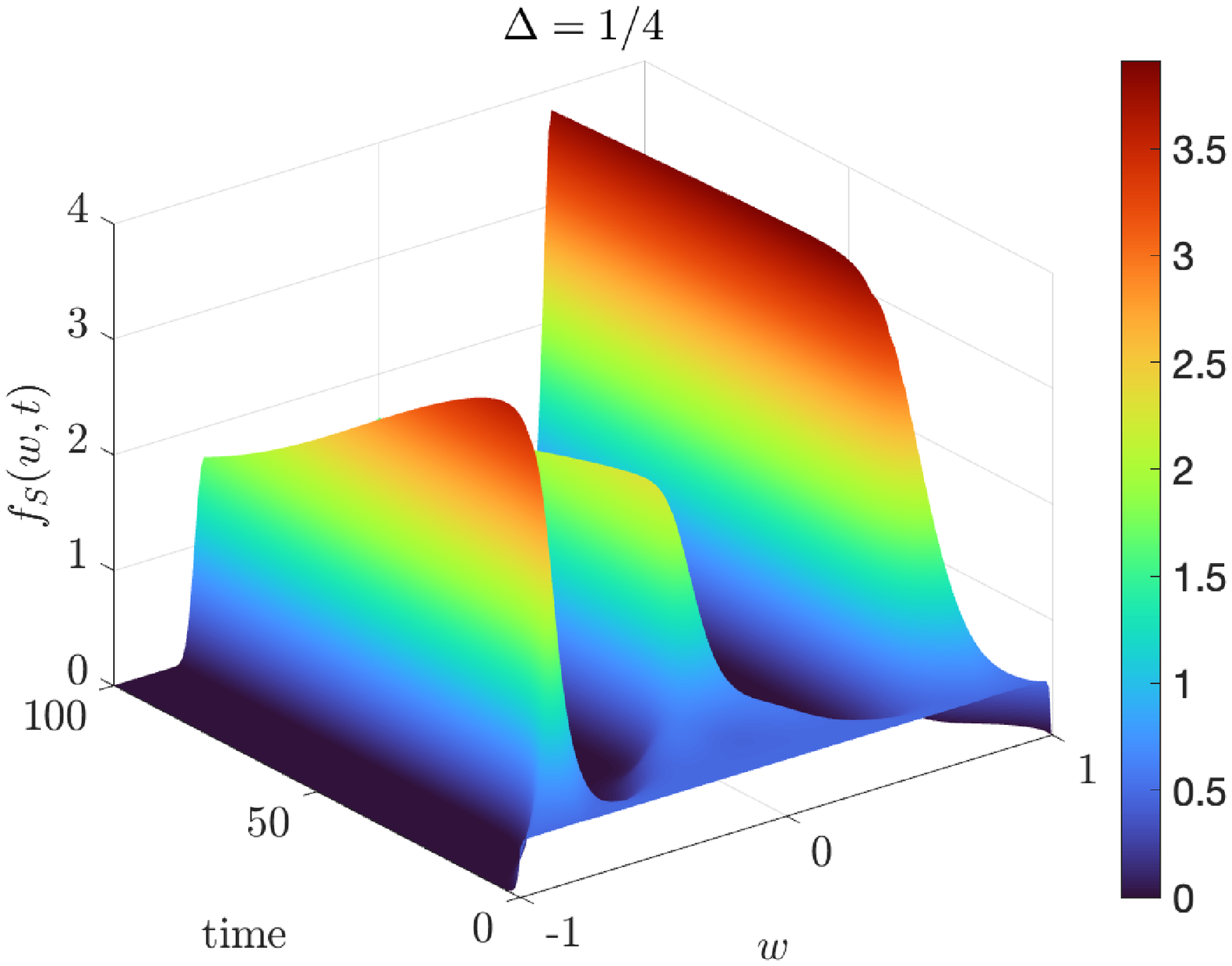}
\includegraphics[scale = 0.3]{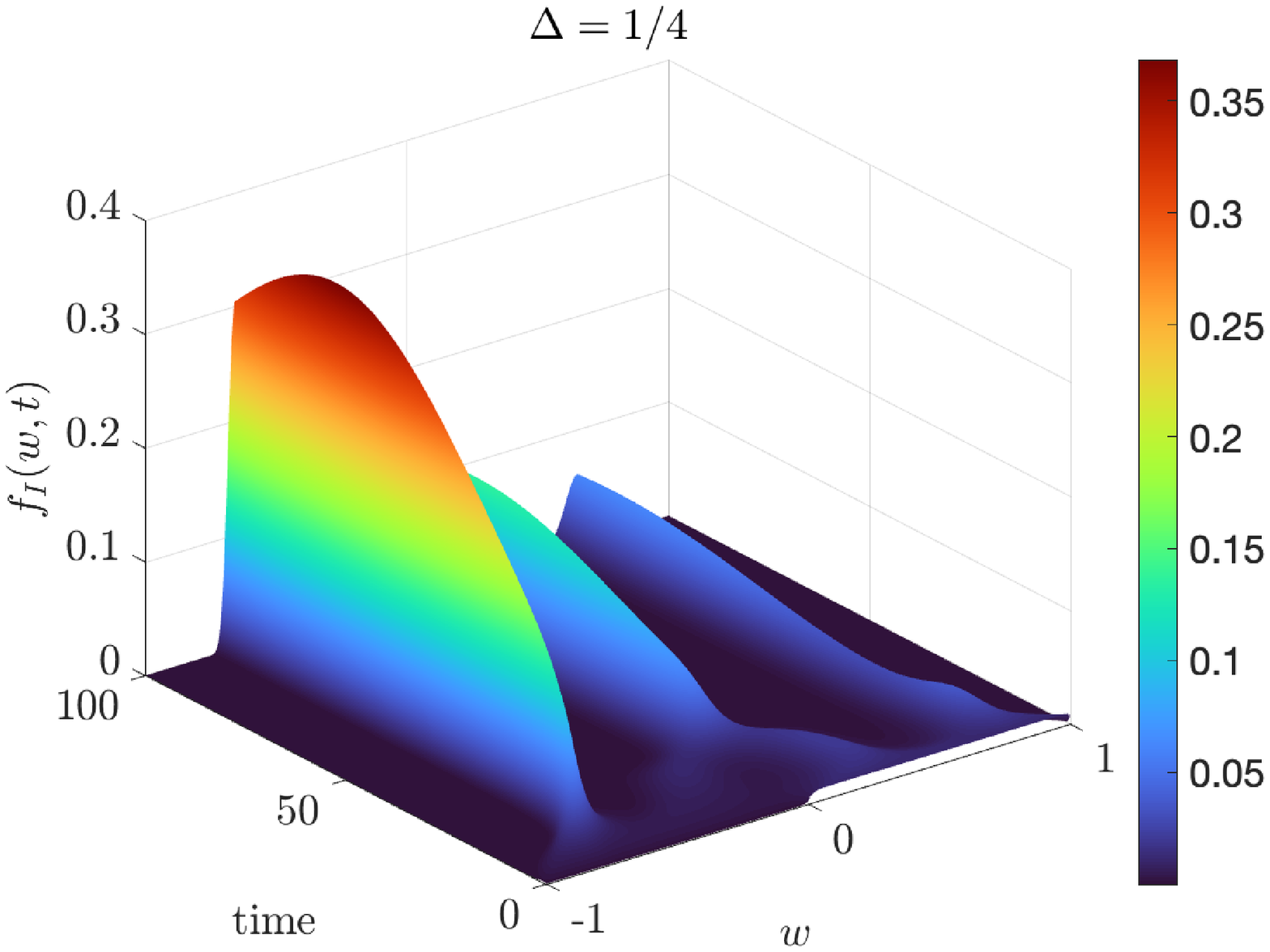}
\caption{\textbf{Test 2b}. We consider a bounded confidence interaction function \eqref{eq:BC} with $\Delta = \frac{1}{4}$. Top row: evolution of mass fractions (left) and mean values (right) for the agents in compartments $\mathcal C$ with $\tau = 1$ and extrapolated from the kinetic model \eqref{eq:kinetic_opinion} with a Fokker-Planck operator $\bar Q(\cdot,\cdot)(w,t)$ of the form \eqref{eq:Q_BC}. Bottom row: evolution of the kinetic distributions for the compartments $S,I \in \mathcal C$. The solution of the Fokker-Planck step \eqref{eq:split_O} has been performed through a semi-implicit SP scheme over the a grid of $N_w = 201$ gridpoints and $\Delta t = 10^{-1}$. Initial distributions defined in  \eqref{eq:init2}.}
\end{figure}

Proceeding as in Section \ref{subsect:FP}, the Fokker-Planck description of a system of agents in the compartment $J \in \mathcal C$ characterized by bounded confidence interactions is given by the following nonlocal operator
\begin{equation}
\label{eq:Q_BC}
\begin{split}
&\bar{Q}_J(f_J,f_J)(w,t) \\
&\quad = \partial_w \left[ \lambda_J\int_{-1}^1 \chi(|w-w_*|\le \Delta)(w-w_*)f_J(w_*,t)dw_* f_J(w,t) \right. \\
&\qquad\left. + \dfrac{\sigma^2_J}{2}\partial_w (D^2(w)f_J(w,t))\right]
\end{split}
\end{equation}
cf. Remark \ref{rem:FP}. The equilibrium distribution of the corresponding nonlocal model is not explicitly computable and the resulting macroscopic models for the evolution of observable quantities may deviate from the ones defined in Section \ref{sect:macro}. Let us consider the densities
\[
g(w) = 
\begin{cases}
\frac{1}{2} & w \in [-1,1] \\
0 & \textrm{elsewhere},
\end{cases}
\qquad 
h(w) = 
\begin{cases}
1& w \in [0,1] \\
0 & \textrm{elsewhere}
\end{cases}
\]
and we consider the initial distributions
\begin{equation}
\label{eq:init2}
\begin{split}
f_S(w,0) &= \rho_S(0)g(w),\qquad f_E(w,0) = \rho_E(0)g(w),\\
f_I(w,0) &= \rho_I(0)h(w),\qquad f_R(w,0) = \rho_R(0)h(w)
\end{split}
\end{equation}
with $\rho_E(0) = 0.01$, $\rho_I(0) = 0.01$, $\rho_R(0) = 0.01$ and $\rho_S(0) = 1-\rho_E(0)-\rho_I(0)-\rho_S(0)$.


In Figure \ref{fig:BC12} we show the evolution of the kinetic distributions $f_S(w,t)$ and $f_I(w,t)$, $t \in [0,100]$ determined by bounded confidence interactions described by the nonlocal Fokker-Planck-type operator \eqref{eq:Q_BC}, with $\Delta = \frac{1}{2}$, $\lambda_J = 1$, and $\sigma^2_J = 10^{-3}$ for all $J \in \mathcal C$. We may observe how the opinion dynamics lead to two separate clusters centered in $-0.5$ and in $0.5$. Furthermore, coherently with the modelling assumptions characterizing the incidence rate $K(f_S,f_I)(w,t)$ in \eqref{eq:K_def}-\eqref{eq:kappa_def},  the cluster with negative opinions looses mass since it is linked to agents with weak protective behaviour. The infection is therefore propagated to these agents and the kinetic distribution $f_I(w,0)$ gains mass for $w<0$. We highlight how the approximated equilibrium density is not coherent with a Beta distribution. Therefore the  evolution of the macroscopic quantities cannot be obtained through a classical closure method and we need to solve the full kinetic model. 

\rev{
\subsection{Test 2c: infection-driven bounded confidence model}
We consider in the nonlocal operator \eqref{eq:Q_BC} the case in which the interaction function depends on the fraction of infected cases $\rho_I(t)$. To this end, we consider the bounded confidence function
\begin{equation}
\label{eq:PBCI}
P(w,w_*) = \lambda_J \chi(|w-w_*|\le \Delta (\rho_I)),
\end{equation}
where $\Delta(\rho_I)$ is a dynamical confidence threshold depending on the epidemic. We further assume that consensus emerges for sufficiently high values of $\rho_I$, mimicking the fact the adoption of a protective behaviour is triggered by the evolution of the epidemic. In particular, we consider
\begin{equation}
\label{eq:DI}
\Delta(\rho_I) = 
\begin{cases}
\Delta_1 & \rho_I \le C_I \\
\Delta_2  & \rho_I >C_I, 
\end{cases}
\end{equation}
with $\Delta_1< \Delta_2 \in [0,2]$. Therefore, opinion clustering is expected if $\rho_I\le C_I$ and consensus if $\rho_I>C_I$. In Figure \ref{fig:test2c_1} we show the evolution of $\rho_I(t)$ and $\rho_R(t)$ in the case of bounded confidence interactions with infection-driven threshold. The initial conditions have been defined in \eqref{eq:BC}. In particular, we consider $\lambda_J = 1$, $\sigma^2_J = 10^{-3}$ and $\Delta_1 = \frac{1}{10}$ and $\Delta_2 = \frac{1}{2}$, so that that the compromise propensity is higher once the cases escalate. To understand the impact of the parameter $C_I$ we consider $C_I = K \cdot 10^{-2}$ with $K = 1,\frac{5}{2},5$. We may observe how the epidemic peak is reduced for small values of $C_I>0$.   At the same time, the number of recovered agents is reduced for small $C_I>0$.  We report also the evolution of the kinetic density $f_S(w,t)$, $t \in [0,200]$ determined by the model \eqref{eq:kin_2} with $\bar Q_J(\cdot,\cdot)$ defined in \eqref{eq:Q_BC} and infection-driven bounded confidence interaction function \eqref{eq:PBCI}. We can observe that the introduced dynamics imply a sharp switch in the compromise process whose effects are also observable the population level. 
\begin{figure}
\centering
\includegraphics[scale = 0.3]{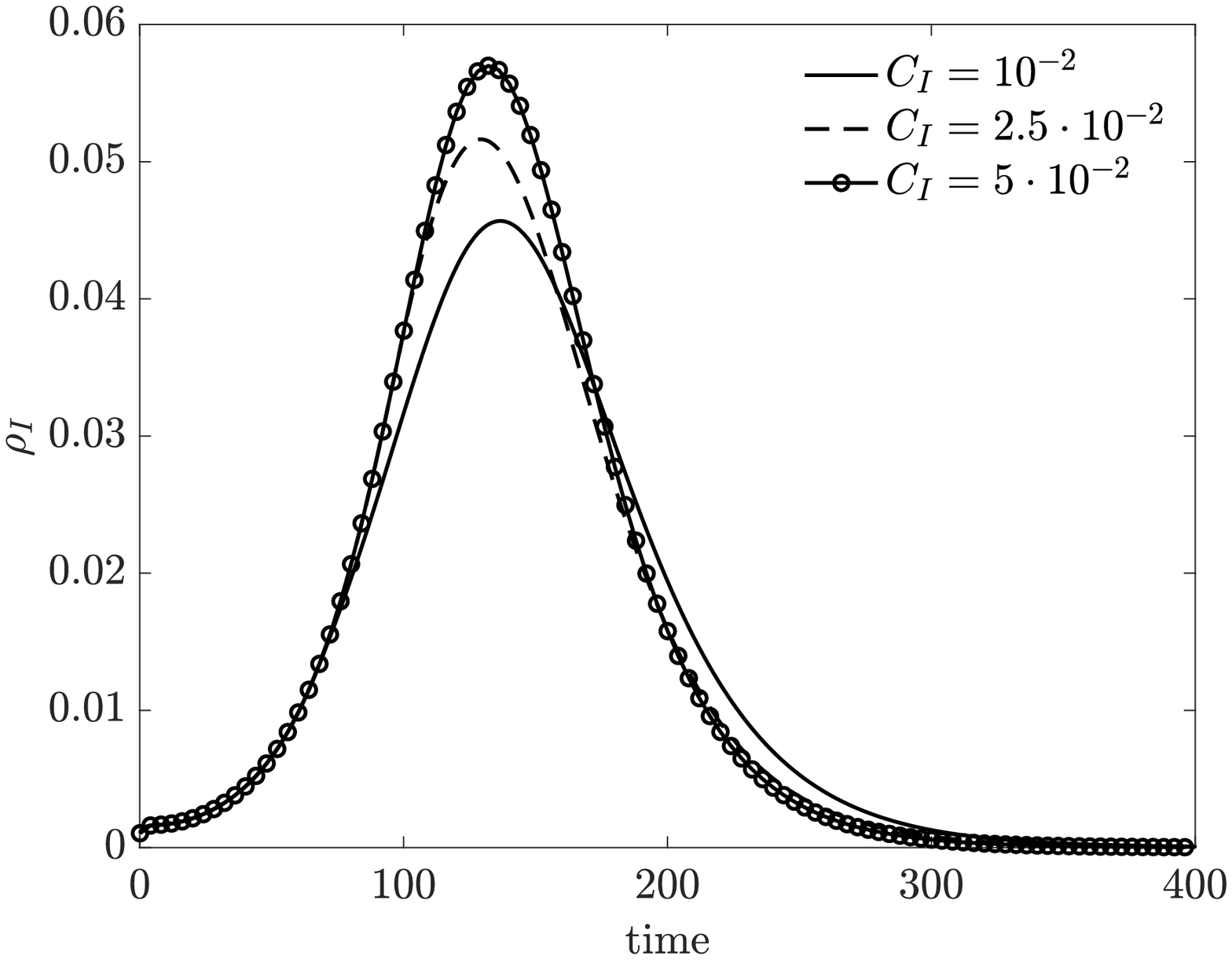}
\includegraphics[scale = 0.3]{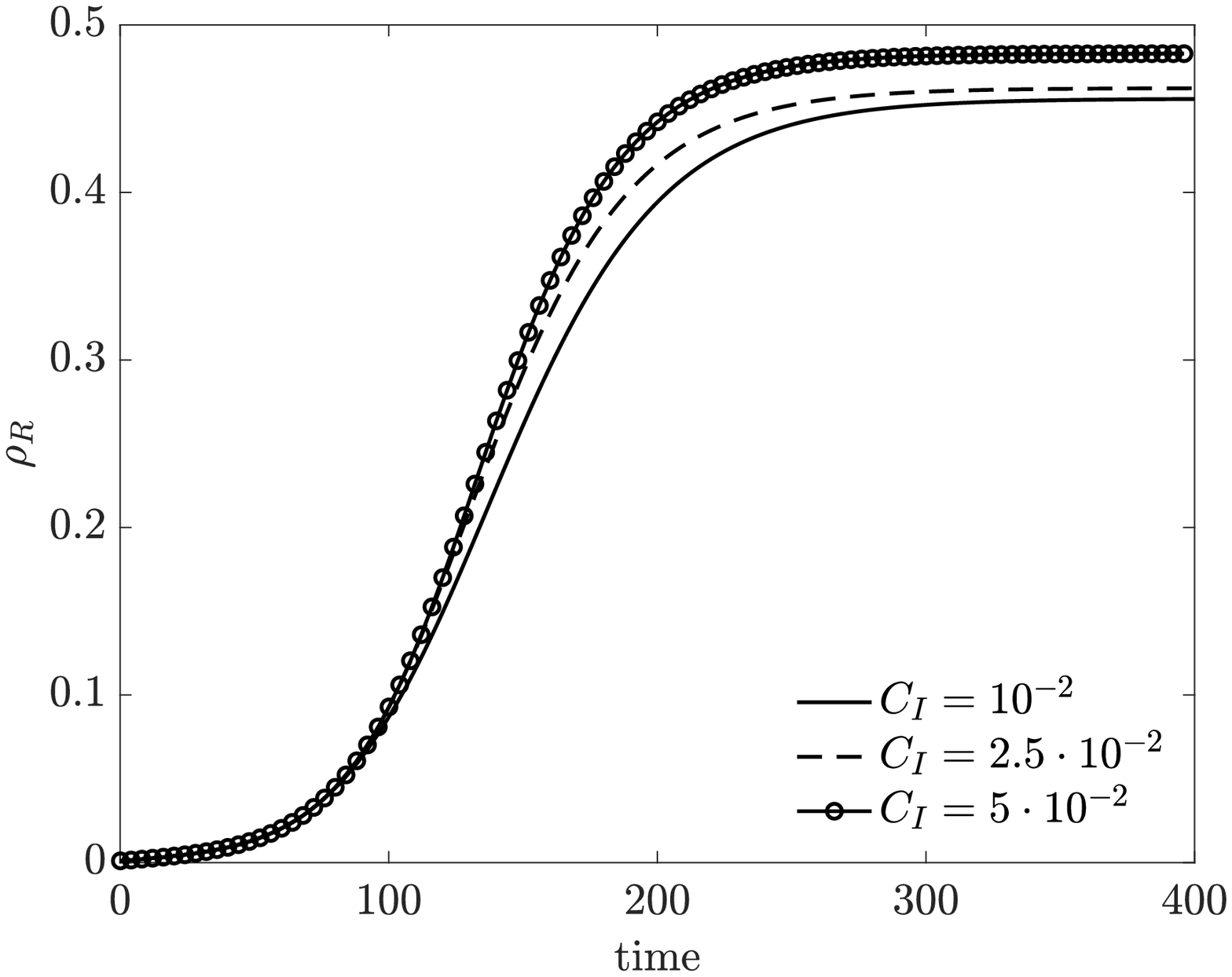} \\
\includegraphics[scale = 0.3]{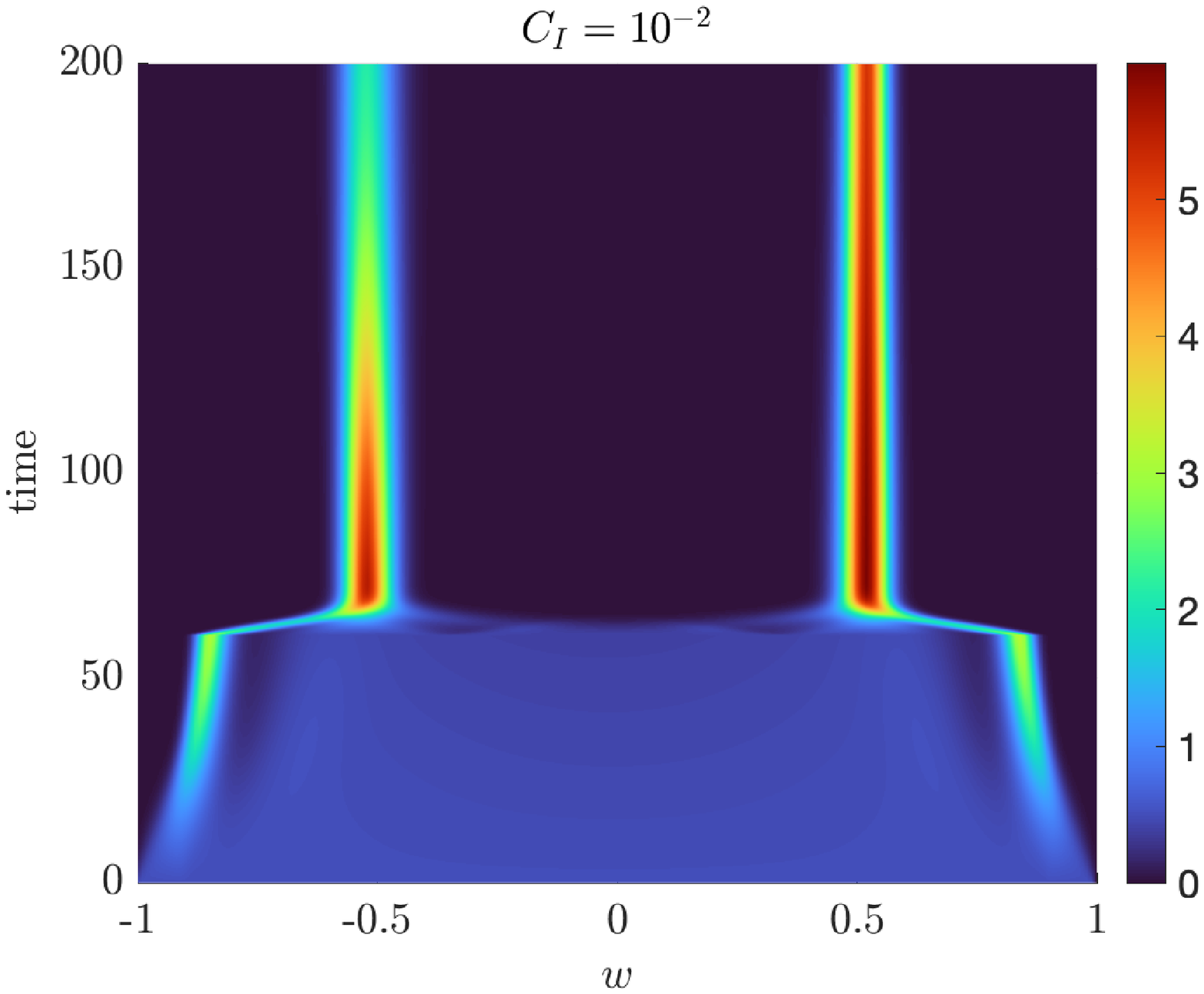}
\includegraphics[scale = 0.3]{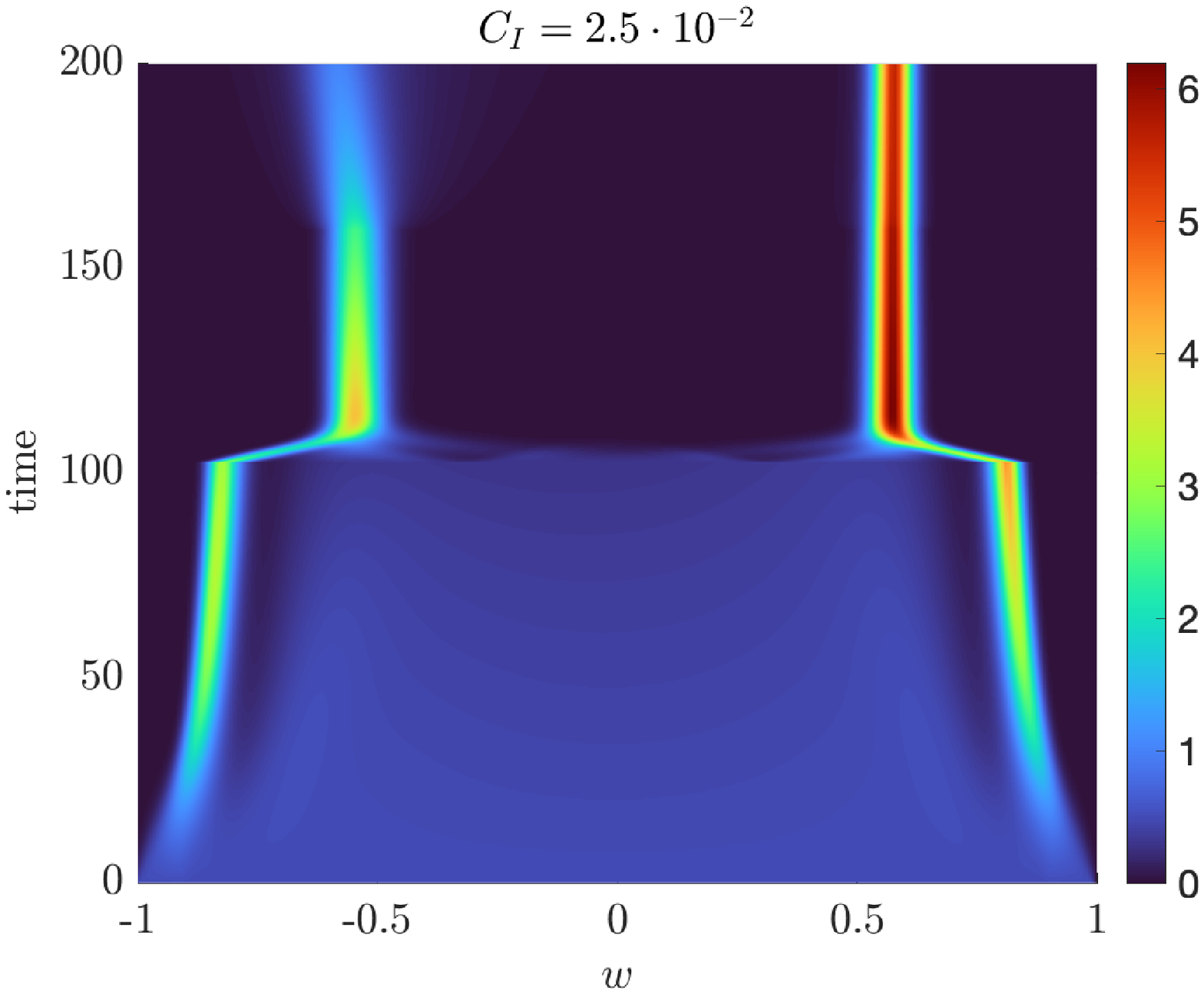}
\caption{\rev{\textbf{Test 2c.} We consider the bounded confidence interaction function \eqref{eq:PBCI} with infection-dependent confidence threshold $\Delta(\rho_I)$ defined in \eqref{eq:DI}. Top row: evolution of $\rho_I(t) = \int_{-1}^1 f_I(w,t)dw$ (left) and $\rho_R(t) = \int_{-1}^1 f_R(w,t)dw$ (right) for several values of $C_I = K\cdot 10^{-2}$ and $K = 1,\frac{5}{2},5$.  Bottom row: evolution of the kinetic distributions for the susceptible compartment in the case $C_I = 10^{-2}$ (left) and $C_I = 5 \cdot 10^{-2}$ (right). The evolution of the kinetic densities has been determined through the semi-implicit SP scheme with $N_w = 201$ gridpoints and $\Delta t = 10^{-1}$. Initial distributions defined in \eqref{eq:init2} with $\rho_I(0) = \rho_E(0) = \rho_R (0) = 10^{-3}$.  } }
\label{fig:test2c_1}
\end{figure}  }

\subsection{Test 3: the impact of opinion polarization on the infection dynamics}

In this test we exploit the derived macroscopic system of mass fractions and mean opinions \eqref{eq:mass}-\eqref{eq:mean} to investigate the relation between opinion polarization and large number of recovered individuals. We recall that, assuming $P\equiv 1$, opinion polarization is observed if $\nu_S>1$, see Section \ref{subsect:FP}. Hence, we consider two main cases, supposing that the mean agents' opinions in all the compartments are exactly alike: the case $m_J(0) = -0.5$, meaning that the agents in each compartment have a bias towards weak protective behaviour, and the case $m_J(0) = 0.5$, meaning that all the agents are biased towards protective behaviour. 

\begin{figure}
\centering
\includegraphics[scale = 0.3]{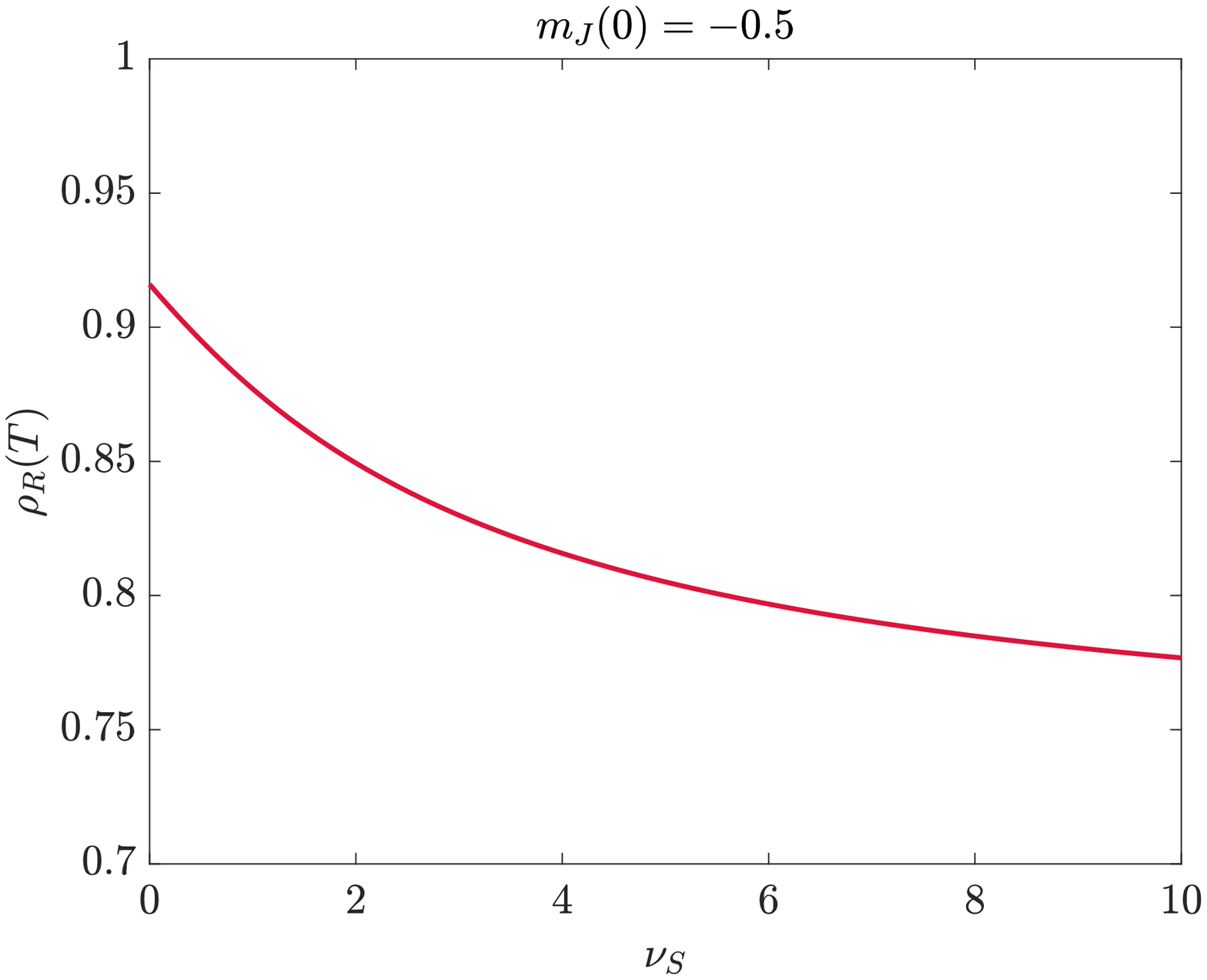}
\includegraphics[scale = 0.3]{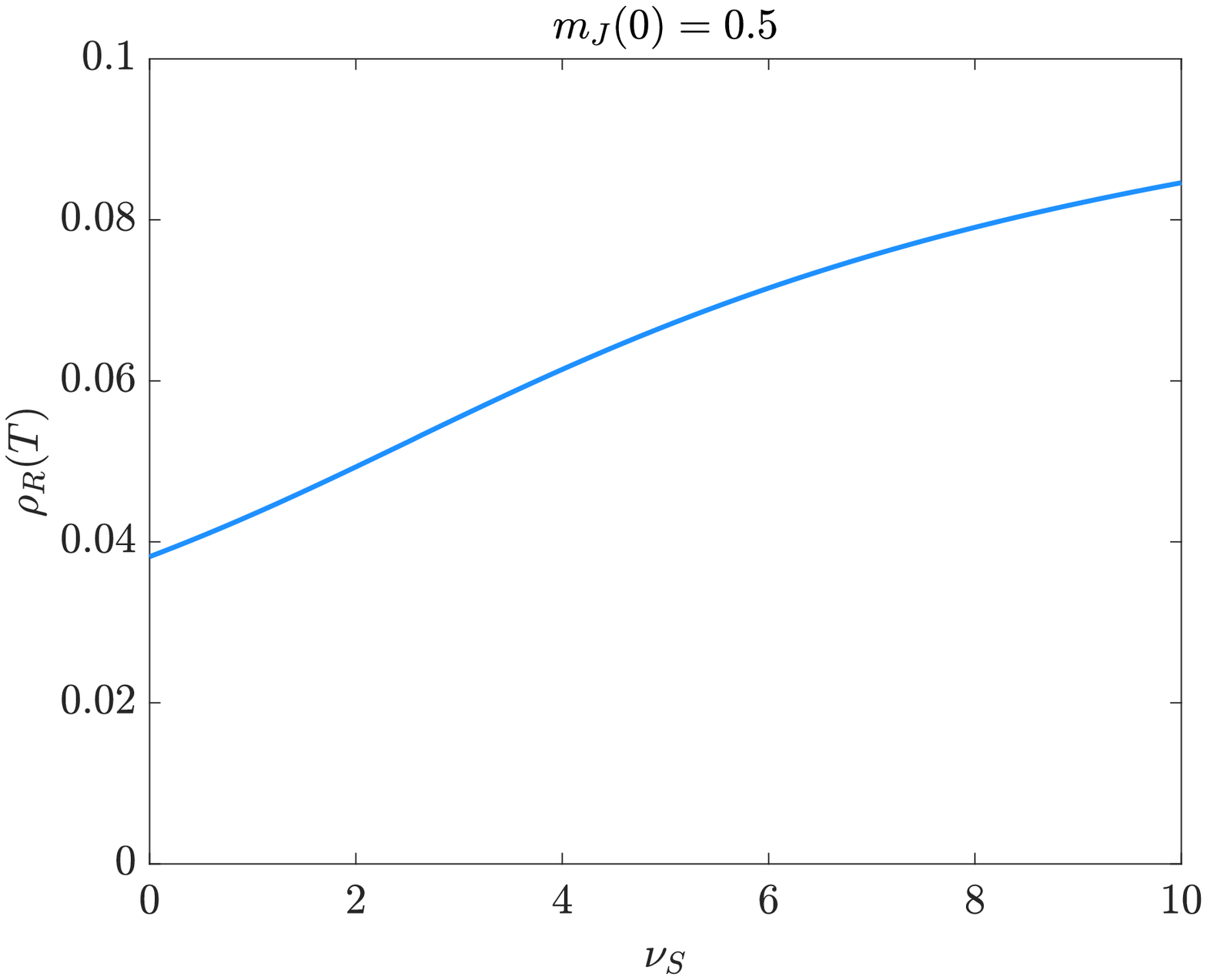}
\caption{\textbf{Test 3}. Impact of the coefficient $\nu_S$ in the large time behaviour of the system \eqref{eq:mass}-\eqref{eq:mean} assuming different initial conditions on the mean opinions of the compartments, $m_J(0) = -0.5$ (left) and $m_J(0) = 0.5$ (right) for all $J \in \mathcal C$. The epidemiological parameters are the same of the previous tests and fixed as follows $\beta = 0.4$, $\sigma_E =1/2$, $\gamma = 1/12$. Furthermore we fixed $\rho_E(0 ) = \rho_I(0) = \rho_R(0) = 0.01$ and $\rho_S(0) = 1-\rho_E(0)-\rho_I(0)-\rho_R(0)$. The system of ODEs is solved through RK4 over a time interval $[0,300]$ with $\Delta t = 10^{-2}$. }
\label{fig:Rinf}
\end{figure}

In Figure \ref{fig:Rinf} we present the large time mass fractions of recovered individuals $\rho_R(T)$ obtained as solution to  \eqref{eq:mass}-\eqref{eq:mean} over the time interval $[0,T]$, $T = 300$, $\Delta t = 10^{-2}$, where we fixed the value $\nu_S \in [0,10]$. In the left figure we consider the case $m_J(0) = -0.5$, whereas in the right figure we consider the case $m_J(0) = 0.5$. We can observe how the effect of opinion polarization strongly depends on the macroscopic initial opinion of the population on protective behaviour.   In details, if the mean opinion is biased towards the adoption of protective behaviour, i.e. $m_J(0)=0.5$, large values of $\nu_S$ trigger an increasing  number of recovered individuals, meaning that the infection have a stronger effect on the society in the presence of polarized opinions. 

On the other hand, if the initial opinion of the population is biased towards the rejection of protective behaviour, i.e. $m_J(0) = -0.5$, opinion polarization is a factor that can dampen the asymptotic number of recovered individuals. Indeed, opinion polarization in this case pushes a fraction of the population towards the two extreme positions and a fraction of agents will stick towards a maximal protective behaviour. 

\section*{Conclusion}
In this work, we considered the effects of opinion polarization on epidemic dynamics. We exploit the formalism of kinetic theory for multiagent system where a compartmentalization of the total number of agents is coupled with their opinion evolution. Kinetic models for opinion formation have been developed in details and are capable to determine minimal conditions for which we can observe polarization of opinions, i.e. the divergence of opinions with respect to a neutral center. Agents' opinions on the adoption of protective behaviour during epidemics is a central aspects for the collective compliance with non-pharmaceutical interventions. Thanks to classical methods of kinetic theory we derived a system of equations that describe the evolution in time of observable quantities that are conserved during the opinion formation process. In particular, considering sufficiently simple  interaction functions and local diffusion functions, we get a second order system of equations for the evolution of mass fractions and mean opinions. This macroscopic system takes into account the social heterogeneities of agents in terms of their opinions and is derived from microscopic dynamics in a SEIR compartmentalization. Thanks to recently developed structure preserving numerical methods, we showed the consistency of the approach by comparing the system of kinetic equations with the set of macroscopic equations.  Furthermore, we analysed more complex interaction functions based on confidence thresholds. The effects of opinion polarization on the asymptotic number of recovered is measured and strongly depends on the initial mean opinion of the population. Indeed, if a positive bias towards protective behaviour is observed opinion polarization is capable to worsen the infection, whereas, if the population tends to reject protective mechanisms, opinion polarization may dampen the total number of infectious agents. Future works will regard more complex opinion formation processes based on leader-follower dynamics and dynamics opinion networks. In future works we will tackle the calibration of the introduced modelling approach \rev{and possible opinion control strategy to prevent the epidemic outbreak.}

\section*{Data availability statement}
The datasets generated during the current study is available from the corresponding author on reasonable request.

\section*{Acknowledgements}
MZ is member of  GNFM (Gruppo Nazionale per la Fisica Matematica) of INdAM, Italy., Italy. MZ acknowledges the support of MUR-PRIN2020 Project No.2020JLWP23 (Integrated Mathematical Approaches to Socio--Epidemiological Dynamics).

\end{document}